%% file: main.tex
\documentclass[article]{elsarticle}
\usepackage[hmargin=3cm]{geometry}
\usepackage{xcolor}
\usepackage{amsmath, amsfonts, amssymb, amsthm, amsbsy}
\usepackage{mathtools, mathrsfs}
\usepackage{bm}
\usepackage{booktabs}
\usepackage{multirow}
\usepackage{placeins}
\usepackage{hyperref}
\usepackage{eso-pic}
\usepackage{url}
\newcommand{\N}{\mathbb{N}}
\newcommand{\R}{\mathbb{R}}
\renewcommand{\vec}[1]{\bm{#1}}
\renewcommand{\d}{\text{d}}

\newdefinition{rmk}{Remark}

\AddToShipoutPictureFG{%
  \ifnum\value{page}=3
    \AtPageLowerLeft{%
      \put(70,75){%
        \makebox(0,0)[bl]{%
          \footnotesize
          \textsuperscript{1}Implementation available at
          \url{https://github.com/sandialabs/SUPN}.
        }%
      }%
    }%
  \fi
}

\renewcommand{\vec}{\bm}

\newtheorem{theorem}{Theorem}
\renewcommand{\d}{\text{d}}

\newtheorem{lemma}{Lemma}
\newtheorem{corollary}{Corollary}

\renewcommand{\d}{\text{d}}

\usepackage[capitalize, nameinlink]{cleveref}
\AddToHook{cmd/appendix/before}{\crefalias{section}{appendix}}
\crefname{section}{Section}{Sections}
\crefname{subsection}{Subsection}{Subsections}
\Crefname{section}{Section}{Sections}
\Crefname{subsection}{Subsection}{Subsections}
\Crefname{figure}{Figure}{Figures}
\crefformat{equation}{\textup{#2(#1)#3}}
\crefrangeformat{equation}{\textup{#3(#1)#4--#5(#2)#6}}
\crefmultiformat{equation}{\textup{#2(#1)#3}}{ and \textup{#2(#1)#3}}
{, \textup{#2(#1)#3}}{, and \textup{#2(#1)#3}}
\crefrangemultiformat{equation}{\textup{#3(#1)#4--#5(#2)#6}}%
{ and \textup{#3(#1)#4--#5(#2)#6}}{, \textup{#3(#1)#4--#5(#2)#6}}{, and \textup{#3(#1)#4--#5(#2)#6}}
\Crefformat{equation}{#2Equation~\textup{(#1)}#3}
\Crefrangeformat{equation}{Equations~\textup{#3(#1)#4--#5(#2)#6}}
\Crefmultiformat{equation}{Equations~\textup{#2(#1)#3}}{ and \textup{#2(#1)#3}}
{, \textup{#2(#1)#3}}{, and \textup{#2(#1)#3}}
\Crefrangemultiformat{equation}{Equations~\textup{#3(#1)#4--#5(#2)#6}}%
{ and \textup{#3(#1)#4--#5(#2)#6}}{, \textup{#3(#1)#4--#5(#2)#6}}{, and \textup{#3(#1)#4--#5(#2)#6}}
\crefdefaultlabelformat{#2\textup{#1}#3}
\crefformat{equation}{(#2#1#3)}
\newcommand{\crefrangeconjunction}{--}
\crefrangeformat{equation}{(#3#1#4)\crefrangeconjunction(#5#2#6)}

\crefname{appendix}{}{}
\Crefname{appendix}{}{}

\usepackage{lineno}

\journal{Journal of Computational Physics}

\begin{document}
\begin{frontmatter}

\title{Expressive Shallow Networks for Physics-Informed Learning:\\ Why Curvature-Aware Optimization and FP64 Are Not Enough}

\author[label1]{Zachary Morrow} 
\author[label2]{Yucheng Fu}
\author[label3]{Pratanu Roy}
\author[label1,label4]{Michael Penwarden}
\affiliation[label1]{organization={Sandia National Laboratories},
            addressline={1515 Eubank Blvd SE}, 
            city={Albuquerque},
            postcode={87123}, 
            state={NM},
            country={USA}}
\affiliation[label2]{organization={Pacific Northwest National Laboratory},
            addressline={3335 Innovation Blvd}, 
            city={Richland},
            postcode={99354}, 
            state={WA},
            country={USA}}
\affiliation[label3]{organization={Lawrence Livermore National Laboratory},
            addressline={7000 East Ave}, 
            city={Livermore},
            postcode={94550}, 
            state={CA},
            country={USA}}
\affiliation[label4]{organization={Kahlert School of Computing, University of Utah}, 
            addressline={50 S. Central Campus Dr.},
            city={Salt Lake City}, 
            postcode={84112},
            state={UT},  
            country={USA}}
\begin{abstract}
Physics-informed neural networks (PINNs) often struggle with high-frequency and high-order partial differential equations (PDEs) due to optimization stiffness and spectral bias, despite possessing sufficient approximation capacity. Existing remedies, including modified feature representations, domain decomposition, higher floating-point precision, and curvature-aware optimizers, can improve convergence but largely act as workarounds for optimization difficulties arising from deep architectures rather than addressing their underlying source. We investigate Physics-Informed Shallow Universal Polynomial Networks (PI-SUPNs), which replace deep nonlinear compositions with learnable polynomial expansions, retaining high approximation capacity within a shallow computational graph. Through comprehensive comparisons with conventional PINNs, physics-informed Kolmogorov--Arnold networks, Fourier-feature PINNs, SIRENs, and finite-basis PINNs, we demonstrate that architectural shallowness can substantially improve trainability on challenging physics. On a high-frequency Helmholtz benchmark, PI-SUPNs are the only architecture to successfully recover the solution, while all competing models fail even under FP64 precision and curvature-aware optimization. However, experiments on higher vibrational modes of a fourth-order Euler–Bernoulli beam reveal that architectural improvements alone are insufficient: accurate solutions require the combination of PI-SUPNs, FP64 precision, and curvature-aware optimization. Additionally, Allen--Cahn experiments show that PI-SUPNs achieve lower solution and derivative errors at reduced
computational cost, with more favorable loss landscapes
than competing architectures. We also establish quasi-local approximation properties and spline-based convergence bounds for SUPNs. Our results show that shallow network architectures are an important mechanism for improving physics-informed trainability, but accurate solutions to the most challenging PDEs can still require FP64 precision and curvature-aware optimization. 
\end{abstract}


\end{frontmatter}

\input{introduction}

\input{preliminary}

\input{experiments}

\input{summary}

\bibliography{references}

\newpage
\appendix 
\crefalias{section}{appendix}
\crefalias{subsection}{appendix}
\crefalias{subsubsection}{appendix}
\input{appendix}

\end{document}

%% file: introduction.tex
\section{Introduction}
Physics-informed neural networks (PINNs) provide a flexible framework for solving forward and inverse partial differential equations (PDEs) by incorporating governing equations into the training objective rather than relying exclusively on labeled simulation data \cite{raissi2019physics,karniadakis2021pinnreview}. Since their introduction, PINNs have been successfully applied to challenging problems in fluid mechanics \cite{cai2021physics,MAO2020112789,arzani2021uncovering}, solid mechanics \cite{HU2024112495,WANG2023116184,haghighat2021physics}, among many other domains \cite{wang2021data,stielow2021reconstruction,10.3389/fphy.2020.00042}.
The appeal of PINNs comes from mesh-free discretization, straightforward incorporation of weakly-enforced physical constraints, and the ability to unify simulation, physics, and sparse experimental data within a single optimization framework.

Despite this promise, PINNs remain difficult to train reliably on challenging PDEs. A growing body of work has identified optimization pathologies arising from gradient imbalance, poor conditioning, spectral bias, and sensitivity to initialization, particularly for multiscale, oscillatory, and stiff nonlinear systems \cite{krishnapriyan2021failure,penwarden2023unified,wang2021gradient,wang2022pinns,rathore2024challenges}. High-order PDEs present an additional challenge because repeated automatic differentiation through deep computational graphs substantially increases computational cost, magnifies floating-point roundoff error, and worsens optimization conditioning \cite{NEURIPS2024_00532321}. Numerous solutions have been proposed, including adaptive loss balancing \cite{mcclenny2020selfadaptive,xiang2022selfadaptive}, curriculum and causal training strategies \cite{wang2024respecting,penwarden2023unified,roy2024exact}, domain-decomposition methods \cite{moseley2023finite,JAGTAP2020113028}, modified implicit representations such as Fourier features and sinusoidal activations \cite{tancik2020fourier,sitzmann2020implicit,penwarden2024kolmogorov}, and improved curvature-aware optimization strategies \cite{kiyani2025optimizing,jnini2026curvature}. However, many of these approaches are workarounds to resolve the inherent ill-conditioning of the optimization problem and preserve the underlying deep architecture responsible for repeated differentiation overhead.

In this work, we investigate Physics-Informed Shallow Universal Polynomial Networks (PI-SUPNs), building on the recently introduced SUPN architecture \cite{morrow2026supn}. SUPNs replace deep nonlinear compositions with learnable tanh-activated polynomial expansions, retaining high approximation capacity within a shallow computational graph. We hypothesize that this architectural shallowness reduces optimization stiffness and repeated differentiation costs, allowing physics-informed training to more effectively realize the network's approximation capabilities. Chebyshev recurrence relations further enable efficient evaluation of the polynomial basis and its derivatives. We also establish quasi-local approximation and convergence results that characterize the ability of SUPNs to approximate localized functions using globally supported polynomial bases.

We evaluate PI-SUPNs against conventional PINNs, physics-informed KANs, Fourier-feature PINNs, SIRENs, and finite-basis PINNs across high-frequency and high-order PDE benchmarks. PI-SUPNs are the only architecture to successfully reconstruct the challenging Helmholtz acoustic field, even when competing methods use 64-bit floating-point precision (FP64) and self-scaling BFGS (SS-BFGS) for optimization. Controlled Allen--Cahn experiments examine accuracy, computational cost, derivative approximation, and loss landscapes. 
For the higher vibrational modes of an Euler--Bernoulli beam,
accurate training is achieved only by combining PI-SUPNs
with FP64 precision and SS-BFGS optimization, highlighting the interplay of architecture, numerical precision, and optimizer choice. The remainder of this paper is organized as follows. Section~2 introduces PI-SUPNs, their approximation properties, baseline architectures, and optimization methods. Section~3 presents the numerical experiments, and Section~4 summarizes the findings.

%% file: preliminary.tex
\section{Preliminary}
Neural networks struggle to fit high-frequency \cite{zhang2025shallow,khodakarami2026spectral,xu2025understanding} and high-order \cite{NEURIPS2024_00532321} PDE solutions. This section introduces the PI-SUPN architecture and its theoretical approximation properties.  We then review baseline architectures for mitigating spectral bias and optimization difficulties, including MLP-based PINNs, PIKANs, Fourier features, SIRENs, and finite-basis PINNs. Finally, we discuss curvature-aware optimization methods and evaluation of generalization under discretization shift and mixed data-physics training.

\subsection{Physics-Informed SUPN}
PINNs, commonly using a multi-layer perceptron (MLP), or more recently a Kolmogorov--Arnold Network (KAN) \cite{SHUKLA2024117290,toscano2025pikans} representation, have successfully been applied to numerous applications in physics \cite{hao2024pinnacle,takamoto2022pdebench}. Shallow Universal Polynomial Networks (SUPNs)\textsuperscript{1} were recently introduced as a shallow but expressive alternative to conventional deep networks for function approximation \cite{morrow2026supn}. Unlike MLPs and KANs, whose expressivity is derived from depth, SUPNs replace all but the last layer as polynomials with learnable coefficients, providing high approximation capacity while maintaining a shallow computational graph written as follows:
\begin{align}
f(\boldsymbol{x}) = \sum_{i=1}^{N} c_i \, \sigma\left( \sum_{j=0}^{M} a_{i,j}T_j(\boldsymbol{x}) \right)
\end{align}
where $\sigma$ is chosen as $\tanh$ and $T_j$ are Chebyshev polynomials of order $j$. The original SUPN work demonstrated strong approximation accuracy, parameter efficiency, and reduced training cost relative to deeper architectures in data-driven function approximation settings.

For physics-informed learning, this architectural distinction is particularly important because PDE training requires repeated differentiation of the network output with respect to the input coordinates. In deep architectures, repeated automatic differentiation introduces substantial computational overhead and increasingly ill-conditioned optimization dynamics due to repeated chain-rule propagation through many layers \cite{NEURIPS2024_00532321,rathore2024challenges}. To improve the efficiency of derivative computations for PI-SUPN training, we precompute Chebyshev derivatives using three-term recurrence relations as described in \ref{app:efficent_polys}.

\rmk{Unlike shallow MLP-based physics-informed architectures \cite{hu2022discontinuity,hu2024shallow} , SUPNs employ learnable polynomial expansions within each nonlinear basis function, providing greater expressivity than conventional single-hidden-layer MLPs.}

\subsubsection{Approximation Properties of SUPNs}
\label{sec:supn-theory}

The approximation properties of SUPNs have been established in \cite{morrow2026supn}, including convergence in Sobolev norms that account for derivatives of the approximated function. These properties are particularly relevant to physics-informed learning, where the governing differential equations impose requirements on both the solution and its derivatives. Here, we additionally consider the ability of SUPNs to approximate localized functions despite employing globally supported polynomial bases.

\begin{theorem}[Quasi-local SUPN approximation]
\label{thm:spline-interpolant-supn}
Let $\Omega\subset\R$ be a compact interval and let $s_{N,k}$ be an $N$-term spline approximation,
\begin{align}
s_{N,k}(x)
&=\sum_{n=0}^{N-1}w_nB_{n,k}(x),
\end{align}
where each $B_{n,k}$ is a nonzero degree-$k$ B-spline supported on $\Omega_{n,k}\subset\Omega$, with $\Omega\setminus\Omega_{n,k}$ containing a nonempty open interval. For $0\leq r\leq k$, $1\leq p\leq\infty$, and any $\delta>0$, there exists a SUPN
\begin{align}
\hat f_{N,M}(x)
&=\sum_{n=0}^{N-1}w_n\hat f_{1,M}^{n,k}(x)
\end{align}
whose basis functions satisfy
\begin{align}
\left\|\hat f_{1,M}^{n,k}\right\|_{
W^{r,p}(\Omega\setminus\Omega_{n,k})}
&\leq (1+\delta)
\epsilon_M(B_{n,k};W^{r,p}(\Omega)),
\end{align}
and whose approximation error satisfies
\begin{align}
\|s_{N,k}-\hat f_{N,M}\|_{W^{r,p}(\Omega)}
&\leq (1+\delta)
\sum_{n=0}^{N-1}|w_n|
\epsilon_M(B_{n,k};W^{r,p}(\Omega)).
\end{align}
Here, $\epsilon_M(B_{n,k};W^{r,p}(\Omega))$ denotes the best degree-$M$ polynomial approximation error of $B_{n,k}$ in $W^{r,p}(\Omega)$.
\end{theorem}

The quasi-local approximation theorem establishes that SUPNs can approximate localized spline functions using globally supported polynomial bases, while bounding the magnitude of individual basis functions outside their target supports. Thus, SUPNs can approximate localized features without explicit domain decomposition. We further establish a spline-based convergence bound that characterizes the dependence of approximation error on network width and polynomial degree. The proof of the quasi-local approximation theorem is provided in \Cref{app:quasi-local-proof}, while the convergence theorem and its corollary are presented in \Cref{app:supn-spline-convergence}.

\subsection{Physics-Informed KAN (PI-KAN)}
Kolmogorov--Arnold Networks (KANs) were proposed as an alternative to standard MLPs by replacing fixed node-based activation functions with learnable univariate basis functions defined on network edges, improving locality and motivated by the Kolmogorov--Arnold representation theorem \cite{liu2025kan}
\begin{align}
f(x_1,\ldots,x_n) = \sum_{q=1}^{2n+1} \Phi_q\left( \sum_{p=1}^{n} \phi_{q,p}(x_p) \right)
\end{align}
where $\Phi_q$, $\phi_{q,p}$ are represented using learnable univariate B-spline expansions. In data-driven SUPNs \cite{morrow2026supn}, SUPNs were shown to outperform KANs in terms of approximation accuracy, parameter efficiency, and training cost across several benchmark problems. A more detailed architectural comparison between SUPNs and KANs can be found in the original SUPN study. More recently, KANs have been extended to physics-informed learning, commonly referred to as Physics-Informed KANs (PI-KANs) \cite{SHUKLA2024117290,toscano2025pikans}. Physics-informed optimization requires repeated differentiation of the network output, which can significantly amplify the stiffness of the loss landscape for deeper architectures, especially on high-order PDEs.

\subsection{Baseline Methods for Spectral Bias}
Standard multilayer perceptrons (MLPs) learn smooth, low-frequency components before higher-frequency structure, which can impair the learning of oscillatory PDE solutions \cite{xu2019frequency}. From a Neural Tangent Kernel (NTK) perspective \cite{jacot2018neural}, this behavior arises because low-frequency eigenmodes of the NTK typically correspond to larger eigenvalues and therefore converge more rapidly during gradient-based optimization. Recent NTK analyses of PINNs have shown that this eigenvalue decay contributes to poor convergence on multiscale and high-frequency PDEs \cite{wang2021eigenvector,wang2022and,saadat2022neural}.

\subsubsection{Fourier Features}
First introduced in \cite{tancik2020fourier}, Fourier Features lift the spatial coordinates ($\boldsymbol{x})$ into a higher-dimensional space before feeding them into the MLP as defined by the following mapping: 
\begin{align}
    \gamma(\boldsymbol{x}) = [\cos(2 \pi \boldsymbol{B}\boldsymbol{x}), \sin(2 \pi \boldsymbol{B} \boldsymbol{x})]
\end{align}
where $\boldsymbol{B}\in\mathbb{R}^{m \times d}$ is initialized from a Gaussian distribution $\mathcal{N}(0,\sigma^2)$ where $d$ is the dimension of the input space and $m$ is the number of sampled Fourier modes. From a function-approximation perspective, Fourier features provide the network with a fixed sinusoidal basis embedding of the input coordinates, enabling subsequent layers to more easily represent combinations of oscillatory modes across multiple frequencies. Fourier embeddings are particularly attractive for PDE problems with periodic boundary conditions, which get strongly enforced \cite{dong2021method,wang2024respecting}.

\subsubsection{SIREN}
Sinusoidal Representation Networks (SIRENs) \cite{sitzmann2020implicit} replace conventional activation functions such as ReLU or tanh with periodic sinusoidal activations to improve the representation of high-frequency signals and their derivatives. A SIREN layer is defined as follows:
\begin{align}
    \phi(\boldsymbol{x})=\sin(\omega_0(W\boldsymbol{x}+b))
\end{align}
where $W$ and $b$ denote the weights and biases, and $\omega_0$ is a frequency scaling parameter controlling the range of frequencies propagated through the network \cite{belbute-peres2023simple}. The authors of \cite{sitzmann2020implicit} also propose a principled initialization strategy which stabilizes optimization compared to standard Xavier \cite{glorot2010understanding} and He \cite{he2015delving} schemes. A key characteristic of SIRENs for PDEs is their ability to accurately represent higher-order derivatives. Since derivatives of sinusoidal functions remain sinusoidal, repeated automatic differentiation does not suffer from the derivative attenuation commonly encountered with saturating activations such as tanh.

\subsubsection{Finite-Basis PINN}
Finite-Basis Physics-Informed Neural Networks (FB-PINNs) \cite{moseley2023finite} claim to mitigate spectral bias and optimization stiffness by decomposing the computational domain into overlapping subdomains, each represented by a local neural network. By restricting each subnetwork to a subdomain, FB-PINNs effectively reduce the frequencies that must be represented by an individual model, improving convergence on multiscale and high-frequency PDEs. The FB-PINN ansatz can be written as 
\begin{align}
u_\theta(\boldsymbol{x})=\sum_{i=1}^{N} w_i(\boldsymbol{x})\,u_{\theta_i}^{(i)}(\boldsymbol{x})
\end{align}
where $u_{\theta_i}^{(i)}$ is the local network and $w_i$ is its associated window function. Unlike Fourier feature embeddings \cite{tancik2020fourier} and SIRENs \cite{sitzmann2020implicit}, which primarily modify the underlying network, FB-PINNs attempt to address spectral bias through spatial decomposition.

\subsection{Improved Optimization}
Optimization is a major bottleneck in deep learning and particularly PINNs due to their highly non-convex loss landscapes \cite{rathore2024challenges}. While phenomena such as spectral bias are frequently discussed and addressed by modification to the neural implicit representations, the choice of optimizer is often the key contributing factor. In highly overparameterized non-convex systems, poor conditioning and optimization may obscure the underlying approximation capabilities of the model itself \cite{kiyani2025optimizing}, leading to misleading architectural comparisons.

While Adam and L-BFGS remain common in PINNs, recent work \cite{kiyani2025optimizing,jnini2026curvature} demonstrates that curvature-aware and quasi-Newton optimization methods \cite{Al-Baali03092014} can substantially mitigate training failures \cite{krishnapriyan2021characterizing,penwarden2023unified} and outperform conventional optimizers on challenging PINN and PI-KAN benchmarks \cite{SHUKLA2024117290}. These findings suggest that spectral bias and optimization failures in PINNs are optimization limitations rather than intrinsic representational deficiencies alone.

\subsubsection{SOAP}
SOAP (Shampoo \cite{gupta2018shampoo} with Adam in the Preconditioner Eigenbasis) \cite{vyas2024soap} is a scalable second-order-inspired optimizer combining Shampoo-style matrix preconditioning with Adam-based adaptive updates to improve convergence on ill-conditioned objectives. A SOAP iteration can be written as 
\begin{align}
\theta_{t+1} = \theta_t - \eta \, Q_t
\frac{\hat m_t}{\sqrt{\hat v_t}+\epsilon}
Q_t^\top
\end{align}
where $Q_t$ is the eigenbasis of the Shampoo preconditioner and $\hat{m}_t$, $\hat{v}_t$ are Adam-style estimates of the first and second moments in that eigenbasis, and $\eta$ is the learning rate. Recent theoretical work has also established convergence guarantees for SOAP under general orthogonal projection operators \cite{li2026soap}.

\subsubsection{SS-BFGS}
Self-Scaled BFGS (SS-BFGS) is a quasi-Newton optimizer that modifies the standard BFGS update through an adaptive self-scaling parameter. Originally introduced in \cite{oren1974selfscaling}, self-scaling methods have recently been investigated for PINNs and PI-KANs \cite{kiyani2025optimizing}. We use the Scimba implementation for SS-BFGS, which uses a closure-based optimization step with a strong Wolfe line search \cite{scimba2025ssbroyden}. A simplified self-scaled BFGS update can be written as
\begin{align}
H_{t+1} = \tau_t \left(I-\rho_t s_t y_t^\top\right) H_t \left(I-\rho_t y_t s_t^\top\right) + \rho_t s_t s_t^\top ,
\end{align}
where $H_t$ is the inverse-Hessian approximation, $s_t=\theta_{t+1}-\theta_t$, $y_t=\nabla \mathcal{L}(\theta_{t+1})-\nabla \mathcal{L}(\theta_t)$, $\rho_t=(y_t^\top s_t)^{-1}$, and $\tau_t>0$ is the self-scaling parameter. When $\tau_t=1$, this iteration reduces to standard BFGS. We use the Scimba default of $\tau_k=\min\left(1,\frac{-y_k^\top s_k}{\alpha_k\,s_k^\top g_k}\right)$. In the numerical results, we include SS-BFGS to evaluate whether PI-SUPNs retain advantages under improved curvature-aware optimization.

\begin{figure}[b]
    \centering
    \includegraphics[width=0.9\textwidth]{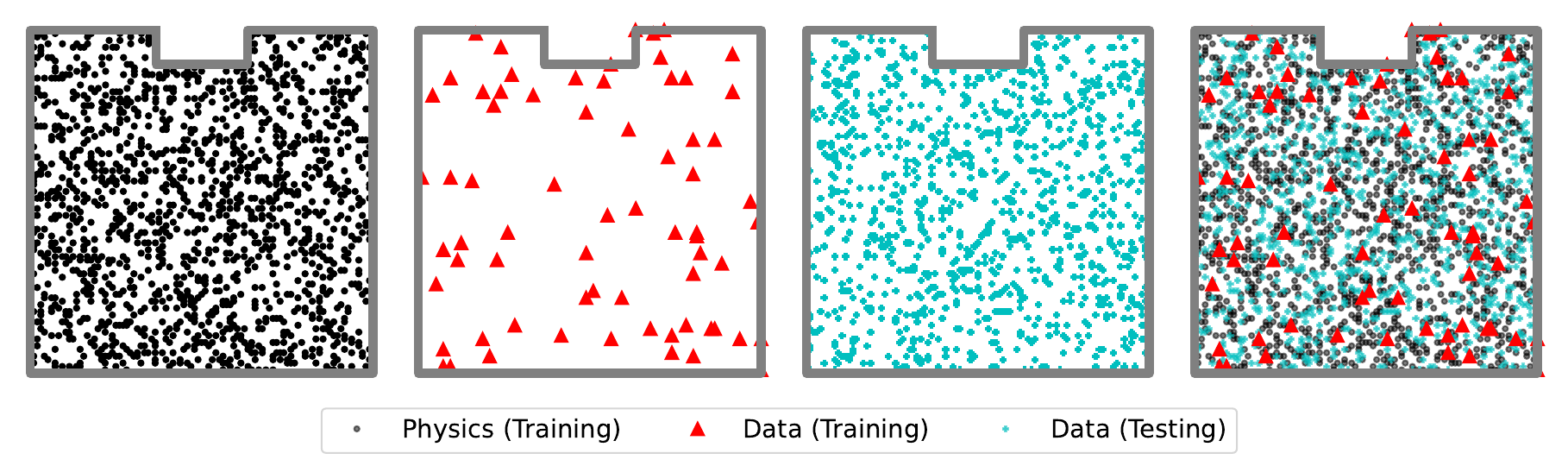}
    \caption{Spatial sampling distributions for mixed data-physics training and super-resolution evaluation, illustrating physics collocation points, training observations, and independent test points.}
    \label{fig:data_fusion}
\end{figure}

\subsection{Discretization Shift and Data-Fusion for Super-Resolution}
\label{sec:data_fusion}
Neural network surrogates are frequently evaluated on the same discretization used during training, potentially obscuring generalization failures. This is particularly important for high-frequency and multiscale PDEs, where sparse collocation points may inadequately resolve oscillations or localized solution features. To assess robustness under discretization shift, we evaluate errors on both training collocation points and independently sampled test discretizations. We also investigate super-resolution through mixed data-physics training, following previous work combining low-resolution observations with high-resolution physics constraints \cite{oommen2025equilibrium}. Figure~\ref{fig:data_fusion} illustrates these different point sets. 
\rmk{Data-fusion training is only performed on the 2D Helmholtz problem. All other problems are purely physics-informed.}

%% file: experiments.tex
\section{Computational Experiments}

We now present several numerical experiments to demonstrate the effectiveness of SUPNs within physics-informed and data-fusion frameworks. \Cref{app:training} describes our training configurations for each problem. We use early stopping at a relative training-loss reduction of $10^{-5}$, and we evaluate errors on independent test sets using the parameters corresponding to the lowest validation loss. We summarize accuracy by taking the median over five random initializations of model weights.

\subsection{Helmholtz Equation}
 \label{sec:helmholtz}
We consider a high-frequency acoustic benchmark motivated by ultrasonic melt processing~\cite{rader2023microstructural,rader2025situ}, modeling the pressure field generated by a vibrating sonotrode immersed in liquid metal. Assuming time-harmonic oscillations, the governing equation reduces to the Helmholtz equation for the complex-valued acoustic pressure amplitude $u(\boldsymbol{x})$:
\begin{equation}
\left( \frac{\partial^2 u}{\partial x^2} + \frac{\partial^2 u}{\partial y^2} \right) + k^2 u = 0 \qquad \text{in } \Omega,
\end{equation}
where $k=\omega/c$ is the wavenumber, $\omega=2\pi f$ is the angular frequency, and $c=1500$ m/s is the speed of sound in the melt.

The computational domain $\Omega$ represents molten metal contained within a crucible. The free surface is modeled as sound-soft, while the side walls are sound-hard:
\begin{align}
u &= 0, && \text{on } \Gamma_{\mathrm{free}}, \nonumber\\
\nabla u \cdot \hat{\boldsymbol{n}} &= 0,
&& \text{on } \Gamma_{\mathrm{wall}},
\end{align}
where $\hat{\boldsymbol{n}}$ is the outward unit normal. Acoustic excitation is imposed through the vibrating sonotrode interface $\Gamma_{\mathrm{probe}}$. For a displacement amplitude $A=0.033$ mm, the normal acceleration amplitude is $a_n=(2\pi f)^2 A$, giving the Neumann boundary condition
\begin{equation}
\frac{\partial u}{\partial y}
=
\rho_0 (2\pi f)^2 A,
\qquad \text{on } \Gamma_{\mathrm{probe}},
\end{equation}
where $\rho_0=1000$ kg/m$^3$.

 \begin{figure}[b!]
    \centering
    \includegraphics[width=0.8\textwidth]{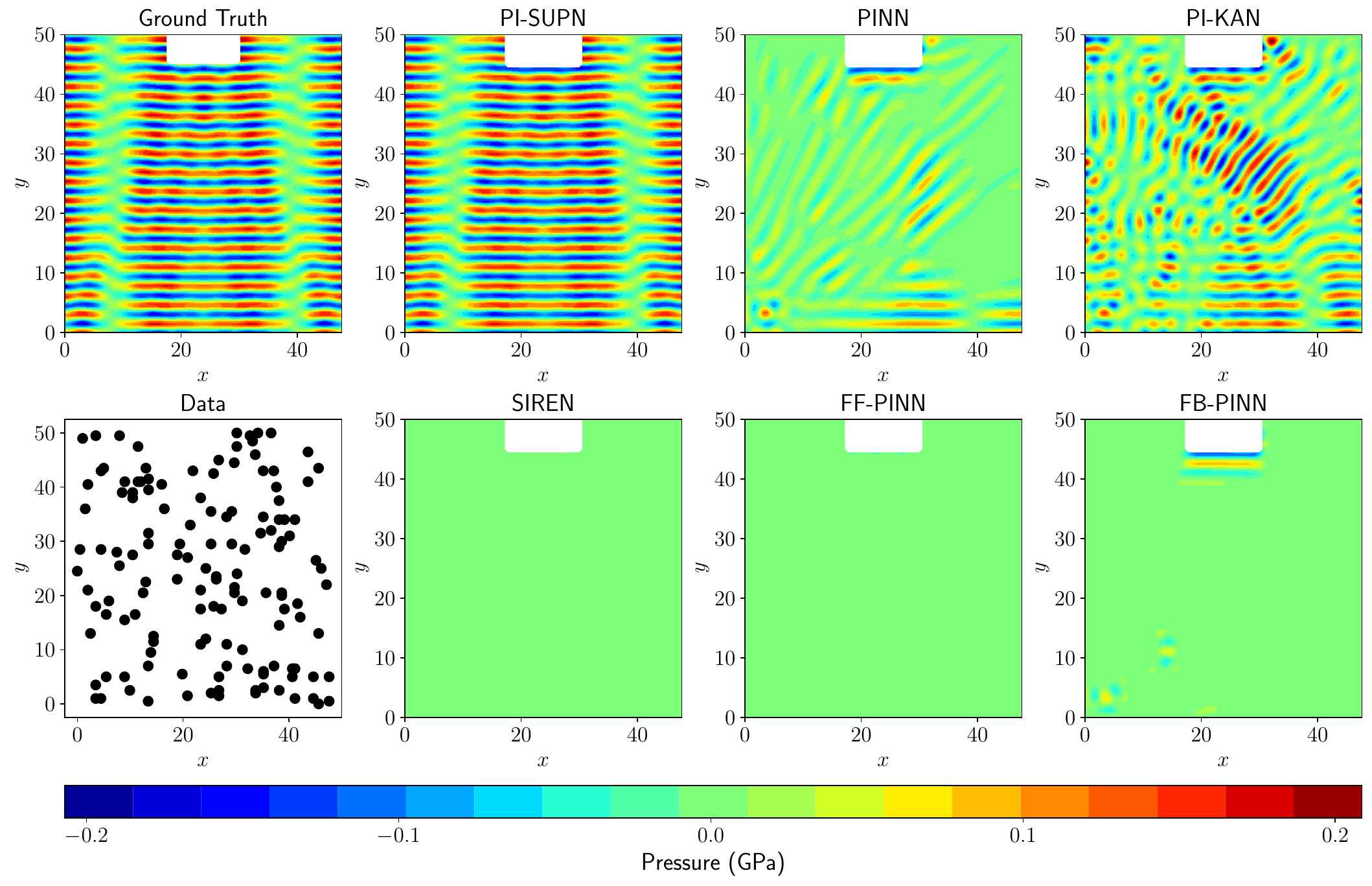}
    \caption{Predicted acoustic pressure fields for the high-frequency Helmholtz problem using Adam + SS-BFGS and FP64. Results show the architecture resulting in the lowest {\em median} relative error over network initialization. The minima over all network initializations are shown in \Cref{app:helmholtz}.}
    \label{fig:helmholtz_4.75E5}
\end{figure}

\begin{figure}[b!]
    \centering
    \includegraphics[width=0.8\textwidth]{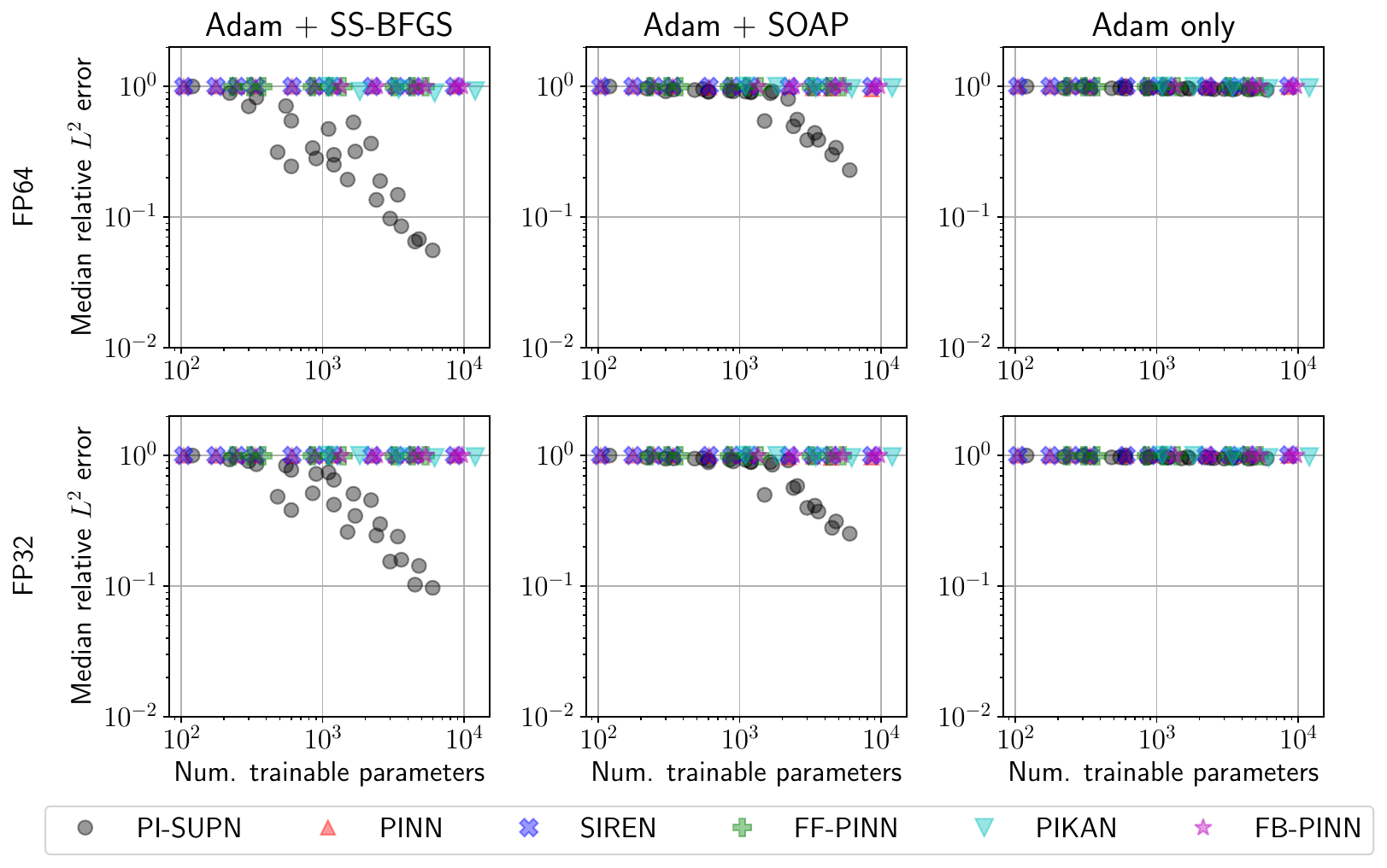}
    \caption{Median relative $L^2$ error over five initializations versus trainable parameter count for the Helmholtz problem with mixed data/physics training. Results compare Adam + SS-BFGS, Adam + SOAP, and Adam-only optimization in FP32 and FP64.}
    \label{fig:helmholtz-convergence}
\end{figure}

We consider an excitation frequency of $f=475$ kHz, producing a highly oscillatory acoustic pressure field. To address the challenging resonant behavior, we employ mixed data-physics training as described in \Cref{sec:data_fusion}, incorporating 136 solution observations from COMSOL~\cite{comsol64}. \Cref{fig:helmholtz_4.75E5} shows that PI-SUPNs successfully reconstruct the oscillatory pressure field, while all competing architectures fail to capture the solution structure. The pointwise errors for the median-selected models are shown in \Cref{fig:helmholtz_pointwise_error}. \Cref{fig:helmholtz-convergence} presents ablation studies across network size, optimizer, and floating-point precision, demonstrating that all five baseline architectures fail to train successfully even with FP64 and SS-BFGS, while PI-SUPNs achieve relative errors on the order of $10^{-2}$. This demonstrates a fundamental difference in trainability between PI-SUPNs and the competing architectures on this challenging high-frequency PDE.

\begin{rmk}
Although the median error models in \Cref{fig:helmholtz-convergence} show that competing architectures generally fail to recover the oscillatory solution, individual initializations can
occasionally achieve better results. However, even when comparing the best-performing models across all initializations, PI-SUPNs retain the lowest prediction error. The corresponding best-model predictions and pointwise errors
are shown in \Cref{fig:helmholtz_4.75E5_min,fig:helmholtz_pointwise_error_min},
respectively, in \Cref{app:helmholtz}.
\end{rmk}

\subsection{Allen--Cahn Equation}
\label{sec:allen-cahn}
We next consider the two-dimensional forced Allen--Cahn equation,
\begin{equation}
\alpha\left(\frac{\partial^2u}{\partial x^2}+\frac{\partial^2u}{\partial y^2}\right)-u(u^2-1)=f(x,y),\qquad (x,y)\in[0,1]^2,
\label{eq:allen-cahn-pde}
\end{equation}
with $\alpha=0.1$ and the manufactured solution
\begin{equation}
u(x,y)=\sin(c_1\pi^2 x)\sin(c_2\pi^2 y)
+\cos(c_2\pi^3 x^2)\cos(c_3\pi^3 y^2)
+\sin(c_3\pi^3 x^3)+\cos(c_4\pi^3 y^3x),
\label{equ:allen_cahn_solution}
\end{equation}
where $f(x,y)$ is determined by substituting the prescribed solution into the PDE. The coefficient vector $\boldsymbol c$ controls the oscillatory structure. We first consider a challenging solution, then deliberately simplify the problem to compare solution accuracy, derivative accuracy, computational cost, and loss landscapes when multiple architectures converge.

\begin{figure}[h!]
    \centering
    \includegraphics[width=0.7\textwidth]{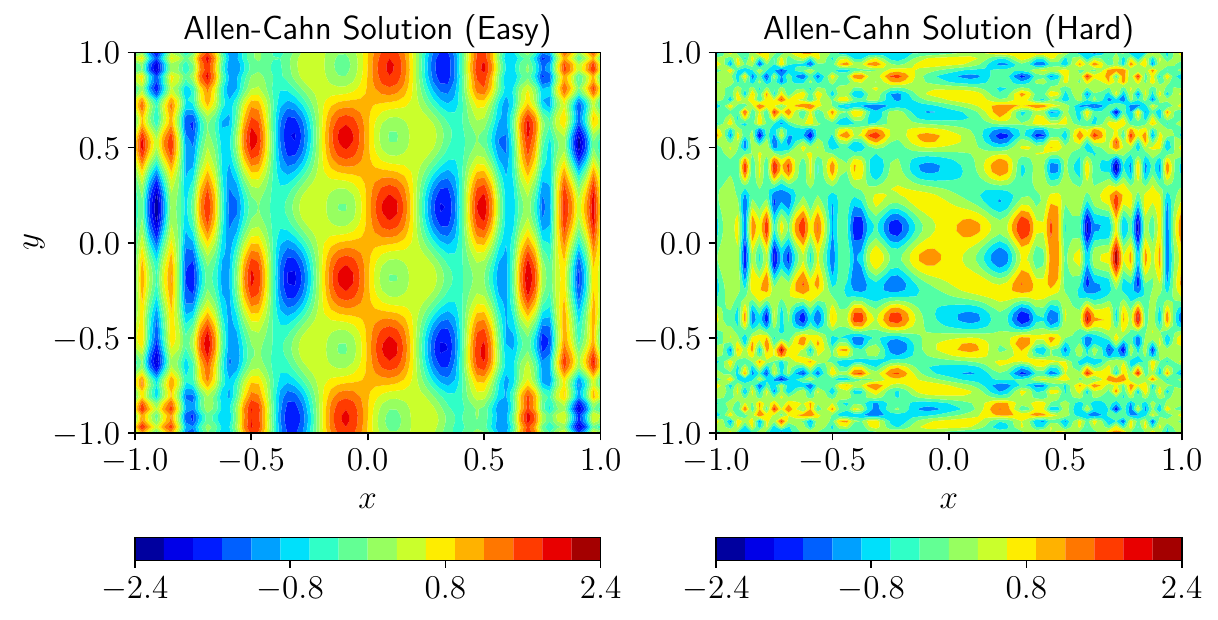}
    \caption{Manufactured solutions for the easy and hard Allen--Cahn benchmarks, illustrating the increased oscillatory structure of the hard case.}
    \label{fig:ac_sols}
\end{figure}

\subsubsection{Hard Allen--Cahn}
\label{sec:allen-cahn-hard}
We choose $\boldsymbol c^{\mathrm{hard}}=(0.681,1.989,0.669,1.614)$, yielding the highly oscillatory solution shown in \Cref{fig:ac_sols}. We compare the architectures over network width and depth, optimizer (Adam + SS-BFGS, Adam + SOAP, and Adam only), and floating-point precision (FP32 and FP64); for FB-PINNs, we additionally vary the number of subdomains.

\begin{figure}[h!]
    \centering
    \includegraphics[width=0.8\textwidth]{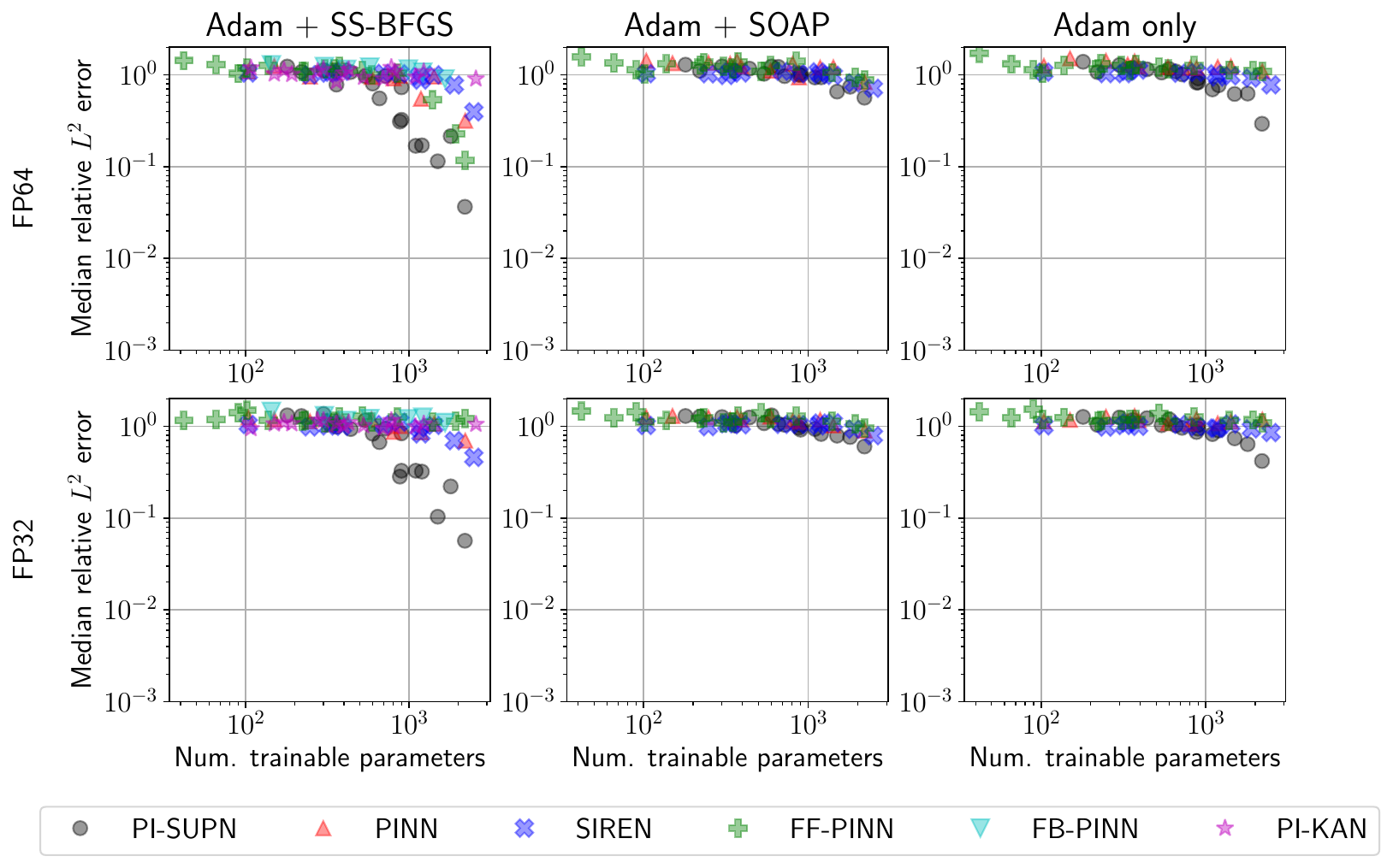}
    \caption{Convergence behavior on the hard Allen--Cahn problem under different combinations of optimizer and floating-point precision (\Cref{sec:allen-cahn-hard}). Reported error is the median over 5 network initializations.}
    \label{fig:ac_hard_convergence}
\end{figure}

\Cref{fig:ac_hard_convergence} shows that PI-SUPNs achieve the lowest errors for many of the tested configurations, particularly in FP32. Performance degrades substantially under SOAP and Adam-only optimization; the full optimizer/precision results are provided in \Cref{tbl:ac_hard_ablation} in the appendix. Because of their computational cost, PI-KANs and FB-PINNs were evaluated only with Adam + SS-BFGS in the hard-case accuracy ablation. Additional per-derivative errors are provided in \Cref{tbl:ac_hard_derivatives} in the appendix.

\subsubsection{Easy Allen--Cahn}

\begin{figure}[h!]
    \centering
    \includegraphics[width=0.8\textwidth]{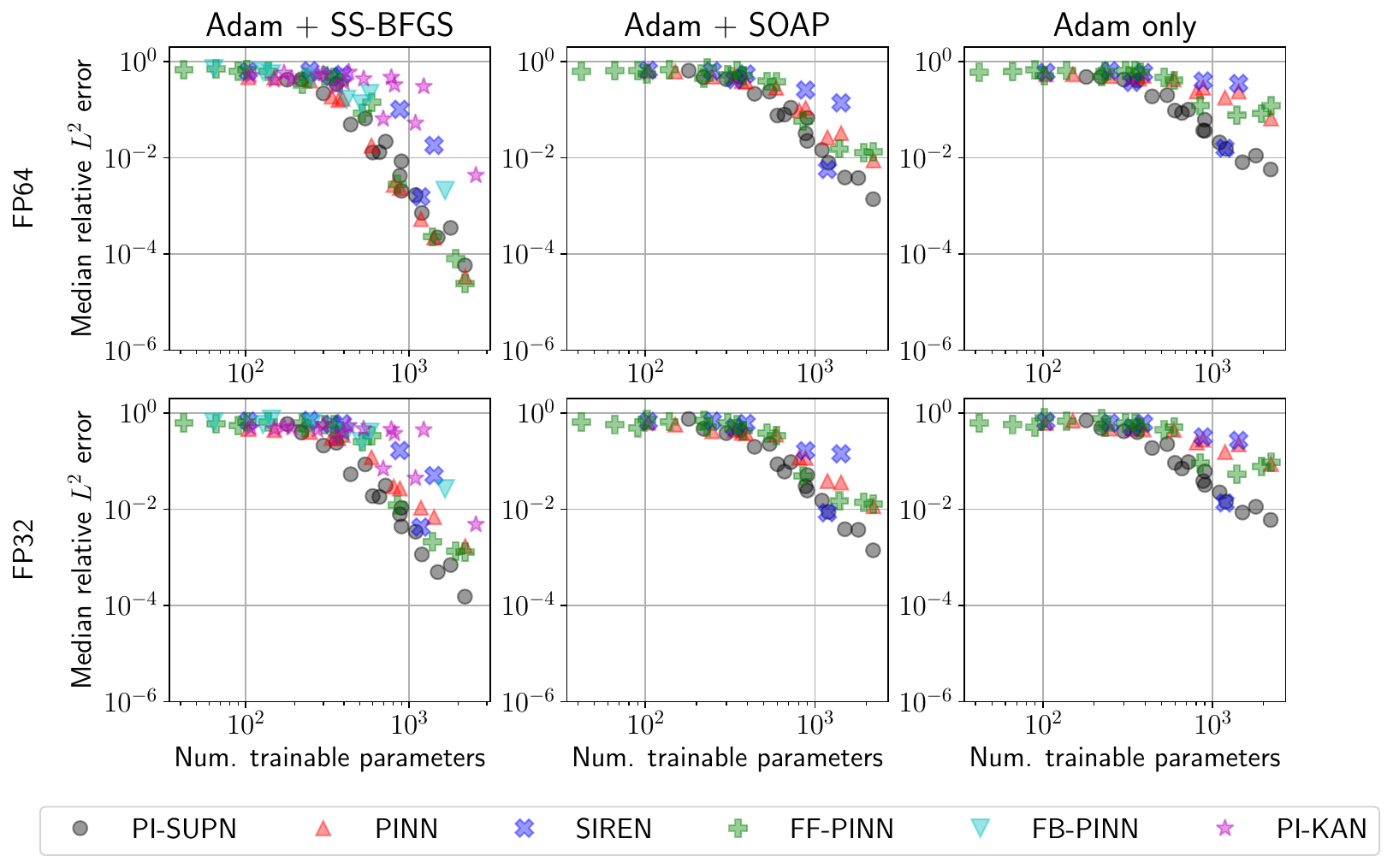}
    \caption{Convergence behavior on the easy Allen--Cahn problem under different combinations of optimizer and floating-point precision (\Cref{sec:allen-cahn-easy}). Reported error is the median over 5 network initializations.}
    \label{fig:ac_easy_convergence}
\end{figure}

\label{sec:allen-cahn-easy}
The preceding Helmholtz and hard Allen--Cahn benchmarks expose substantial differences in trainability, making it difficult to compare optimization characteristics when competing models fail to converge. We therefore consider the deliberately simplified solution with $\boldsymbol c^{\mathrm{easy}}=(1.576,0.857,0.017,0.780)$ (\Cref{fig:ac_sols}), for which several architectures begin approximating the target solution under all optimizer and precision ablations given sufficient network size. 

\begin{figure}[h!]
    \centering
    \includegraphics[width=0.8\textwidth]{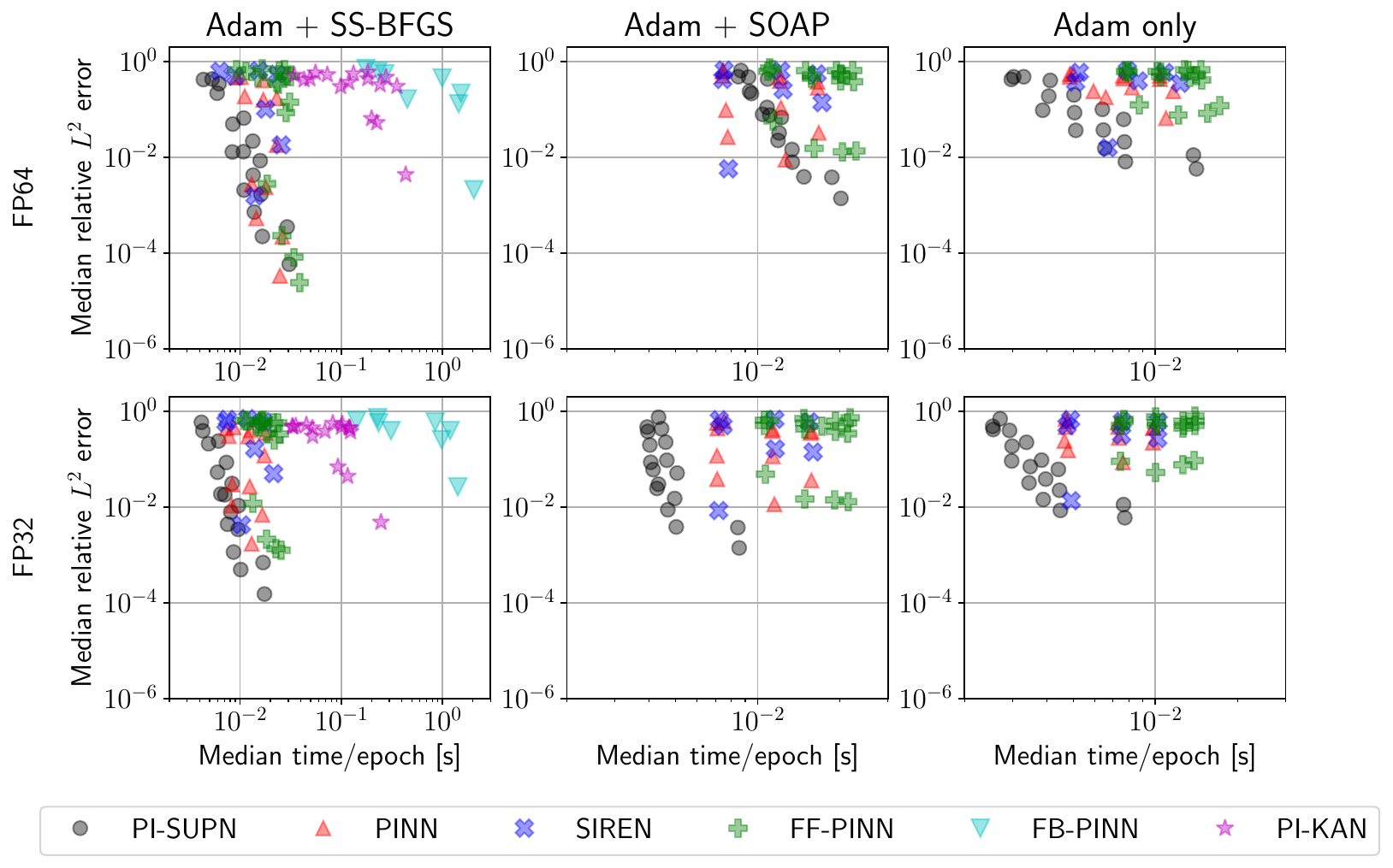}
    \caption{Error vs. runtime for different combinations of optimizer and floating-point precision on the easy Allen--Cahn problem (\Cref{sec:allen-cahn-easy}). Reported error is the median over 5 network initializations.}
    \label{fig:ac_easy_runtime}
\end{figure}

\Cref{fig:ac_easy_convergence} compares accuracy across network sizes, optimization methods, and FP32/FP64 precision. Under FP32, PI-SUPNs attain the lowest error; they also outperform the baselines under Adam + SOAP and Adam-only training. The complete optimizer/precision accuracy table is provided in \Cref{tbl:ac_easy_ablation} in the appendix. \Cref{fig:ac_easy_runtime} shows the corresponding accuracy--runtime tradeoff, with particularly high training costs for PI-KANs and FB-PINNs. This easier Allen-Cahn problem enables meaningful comparison of accuracy and computational cost among successfully trained models.

\begin{table}[h!]
    \centering
    \begin{tabular}{c||c|c|c|c|c}
        \toprule
        Method & $u$ & $u_x$ & $u_y$ & $u_{xx}$ & $u_{yy}$ \\ \hline
        PI-SUPN (ours) & $\mathbf{1.5 \times 10^{-4}}$ & $\mathbf{4.4 \times 10^{-4}}$ & $\mathbf{7.9 \times 10^{-4}}$ & $\mathbf{8.7 \times 10^{-4}}$ & $\mathbf{1.6 \times 10^{-3}}$ \\
        PINN & $1.7 \times 10^{-3}$ & $2.3 \times 10^{-3}$ & $4.5 \times 10^{-3}$ & $3.8 \times 10^{-3}$ & $5.5 \times 10^{-3}$ \\
        SIREN & $2.7 \times 10^{-3}$ & $7.1 \times 10^{-3}$ & $1.2 \times 10^{-2}$ & $1.1 \times 10^{-2}$ & $1.8 \times 10^{-2}$ \\
        FF-PINN & $1.3 \times 10^{-3}$ & $2.5 \times 10^{-3}$ & $4.6 \times 10^{-3}$ & $4.2 \times 10^{-3}$ & $6.4 \times 10^{-3}$ \\
        PI-KAN & $1.8 \times 10^{-3}$ & $3.4 \times 10^{-3}$ & $5.9 \times 10^{-3}$ & $8.7 \times 10^{-3}$ & $1.2 \times 10^{-2}$ \\
        FB-PINN & $2.2 \times 10^{-3}$ & $6.6 \times 10^{-3}$ & $1.1 \times 10^{-2}$ & $1.1 \times 10^{-2}$ & $1.8 \times 10^{-2}$ \\
        \bottomrule 
    \end{tabular}
    \caption{Median $L^2$ errors for each spatial derivative on the easy Allen--Cahn problem (\Cref{sec:allen-cahn-easy}) using Adam + SS-BFGS and FP32.}
    \label{tbl:ac_easy_derivatives}
\end{table}

The derivative comparison in \Cref{tbl:ac_easy_derivatives} evaluates the fitted solutions and their first- and second-order spatial derivatives under FP32 and Adam + SS-BFGS. PI-SUPNs achieve the lowest error for the solution and every reported derivative.

\begin{figure}[h!]
    \centering
    \includegraphics[width=0.8\textwidth]{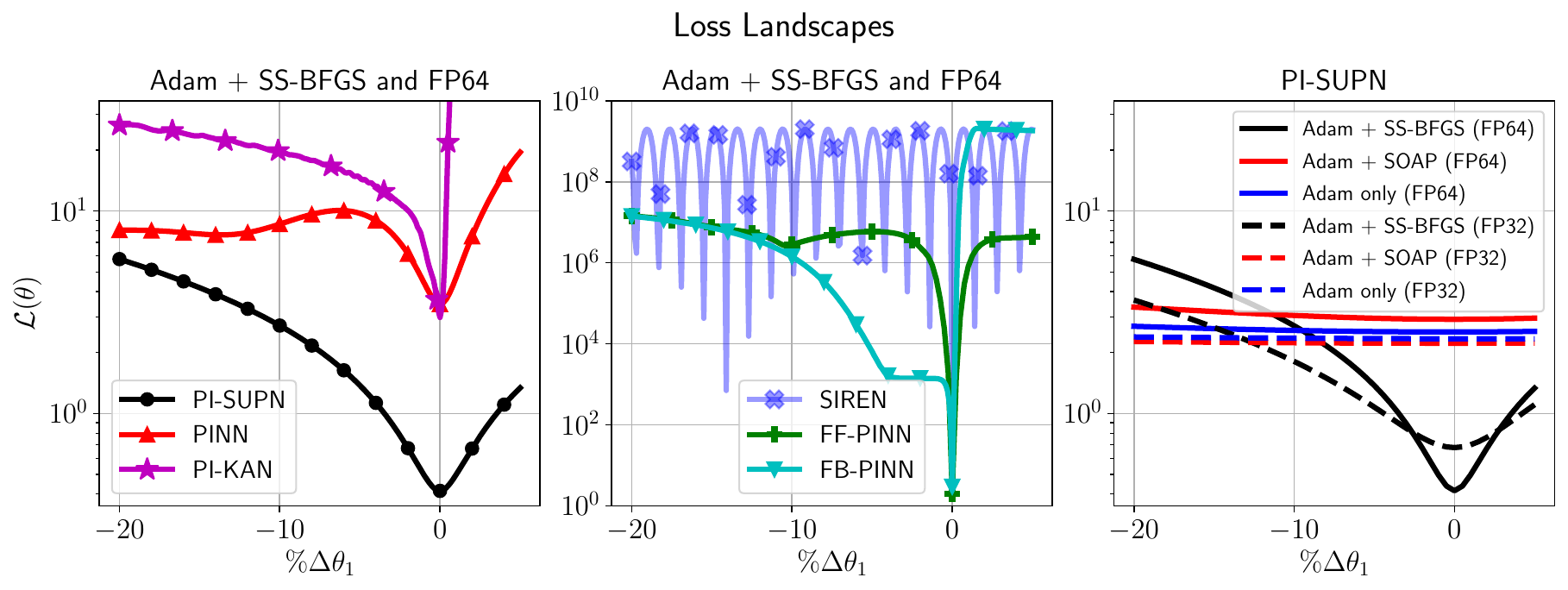}
    \caption{One-dimensional loss-function slices around fitted network parameters for the best-performing models on the easy Allen--Cahn problem. Every perturbed parameter is in the first model layer. Left and center plots use Adam + SS-BFGS and FP64; right plot uses PI-SUPNs.}
    \label{fig:loss_landscapes}
\end{figure}

\paragraph{Loss-landscape analysis} We further compare optimization behavior through one-dimensional loss-landscape slices around fitted model parameters, a diagnostic commonly used to study neural-network optimization \cite{goodfellow2015qualitatively,li2018visualizing,garipov2018loss,draxler2018essentially}. Initialization sensitivity is also relevant to the reliability of physics-informed training \cite{zhang2019quantifying,hao2024stability}. In \Cref{fig:loss_landscapes}, the PI-SUPN slice exhibits a smooth, approximately convex profile over the plotted perturbation interval, whereas several baselines show multiple local minima or a relatively flat region. The right panel compares PI-SUPN loss profiles across optimizers and floating-point precisions. 

\subsection{Euler--Bernoulli Beam}
\label{sec:tuning-fork}

\begin{figure}[h!]
    \centering
    \includegraphics[width=0.8\textwidth]{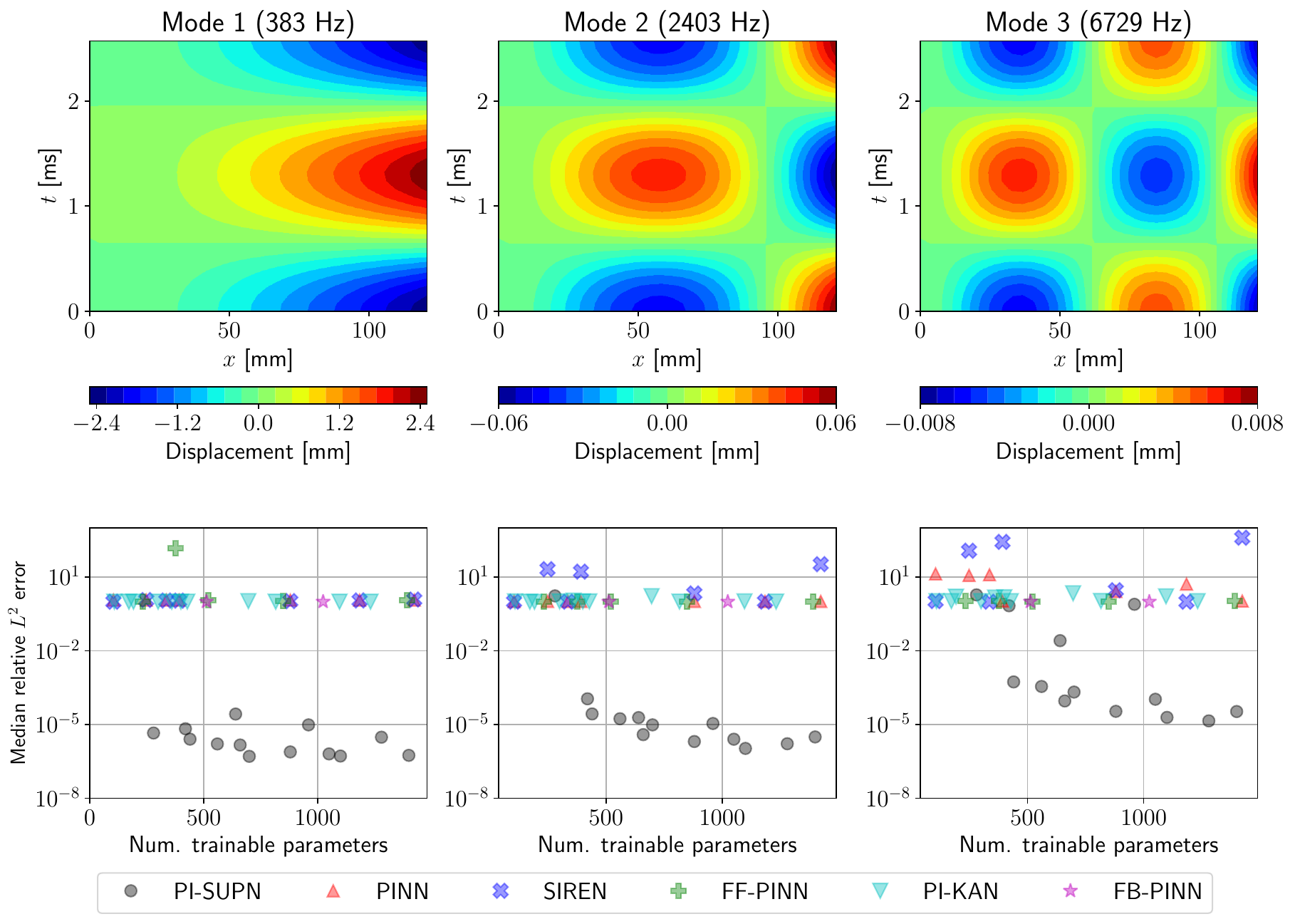}
    \caption{Top: First three vibrational modes of the tuning fork. Bottom: Median relative $L^2$ error over five initializations versus trainable parameter count, using FP64 and Adam + SS-BFGS for each mode.}
    \label{fig:beam-modes}
\end{figure}

We consider the transverse vibrations of a steel tuning-fork prong, modeled by the fourth-order Euler--Bernoulli beam equation,
\begin{equation}
    \frac{\partial^2 u}{\partial t^2} + \frac{EI}{\mu}\frac{\partial^4 u}{\partial x^4} = 0,
    \qquad x\in[0,L],\quad t>0,
    \label{equ:euler-bernoulli}
\end{equation}
where $u(x,t)$ is the displacement, $E=200$ GPa is Young's modulus, $I$ is the second moment of area, and $\mu=\rho A$ is the linear mass density, with $\rho=7850$ kg/m$^3$ and cross-sectional area $A$. The prong is clamped at $x=0$ and free at $x=L$, with initial displacement and zero initial velocity given by
\begin{equation}
\begin{aligned}
    u(0,t)&=0, & u_x(0,t)&=0, & u_{xx}(L,t)&=0, & u_{xxx}(L,t)&=0,\\
    u(x,0)&=-\frac{PL}{EI}\left(\frac{x^2}{2}-\frac{x^3}{6L}\right),
    & u_t(x,0)&=0.
\end{aligned}
\label{equ:euler-bernoulli-bc}
\end{equation}
Following \cite{rossing1992acoustics}, the prong has a $9$ mm $\times$ $7$ mm cross-section and length $L=122$ mm, giving $I\approx2.6\times10^{-10}$ m$^4$ and $\mu\approx0.49$ kg/m.

Separation of variables \cite[pp.~399--401]{meirovitch2001fundamentals} yields the modal expansion
\begin{equation}
    u(x,t)=\sum_{n=1}^{\infty} A_n X_n(x)\cos(\omega_n t)
    \approx\sum_{n=1}^{N} A_n X_n(x)\cos(\omega_n t)\eqqcolon u_N(x,t),
    \label{eq:beam-modal-expansion}
\end{equation}
where
\begin{align}
    X_n(x) &=\cosh(k_nx)-\cos(k_nx)
    +\frac{\cos(k_nL)+\cosh(k_nL)}{\sin(k_nL)+\sinh(k_nL)}
    \bigl(\sin(k_nx)-\sinh(k_nx)\bigr),\nonumber\\
    \omega_n&=k_n^2\sqrt{EI/\mu} \\
    A_n&=\int_0^L u(x,0)X_n(x)\,\mathrm{d}x
    =(-1)^n\frac{2P}{LEI\,k_n^4}\nonumber,
    \label{eq:beam-modes}
\end{align}
and $k_n$ satisfies 
$$
\cos(k_nL)\cosh(k_nL)+1=0, \qquad k_{n} < k_{n+1} \, .
$$
The fundamental frequency is $\omega_1/(2\pi)=383.47$ Hz. We separately train the models on the first three modes over one oscillation period for each mode, with a three-mode truncation error of $9.1\times10^{-4}$ in the reference solution.

\Cref{fig:beam-modes} shows the modal solutions and the corresponding accuracy sweeps under FP64 and Adam + SS-BFGS. PI-SUPNs are the only architecture to achieve convergence to the solution across all three modes. Certain initializations (not the median) of conventional PINNs attain approximately $10^{-3}$ error on the fundamental mode but degrade to $10^{-2}$ on mode 2 and order-unity error on mode 3. SIRENs, Fourier-feature PINNs, PI-KANs, and FB-PINNs remain near unit error across the three modes under the same training conditions. The optimizer and floating-point precision ablations for the fundamental mode are reported in \Cref{tbl:tuning-fork-mode-1}.

\Cref{tbl:tuning-fork-modes-2-and-3} demonstrates the critical interaction between floating-point precision and curvature-aware optimization for the higher vibrational modes. PI-SUPNs achieve relative errors of $1.1\times10^{-6}$ and $1.4\times10^{-5}$ for modes 2 and 3, respectively, only when FP64 precision is combined with SS-BFGS. Switching to FP32 or replacing SS-BFGS with SOAP or Adam causes errors to increase by several orders of magnitude, often exceeding unity. The training-time comparison is provided in \ref{app:beam-additional} (\Cref{tbl:tuning-fork-runtime}). Under FP64 and Adam + SS-BFGS, all three PI-SUPN modal errors are below the three-mode truncation error, so truncation dominates the resulting approximation of the combined solution.

\rmk{These results demonstrate that successful training on challenging high-order PDEs requires the combination of a shallow architecture with favorable optimization characteristics, sufficient floating-point precision, and effective curvature-aware optimization; failure to satisfy any one of these requirements leads to unsuccessful training in the evaluated configurations.}

\begin{table}[h!]
    \centering 
    \begin{tabular}{l|c||c|c|c}
        \toprule
        & Method & Adam + SS-BFGS & Adam + SOAP & Adam only \\ \hline
        \multirow{6}{*}{\rotatebox[origin=c]{90}{FP64}} & PI-SUPN (ours) & $\mathbf{5.1 \times 10^{-7}}$ & $\mathbf{1.3 \times 10^{-3}}$ & $\mathbf{5.2 \times 10^{-1}}$ \\
        & PINN & $1.2 \times 10^{0}$ & $8.5 \times 10^{-1}$ & $1.2 \times 10^{0}$ \\
        & SIREN & $1.0 \times 10^{0}$ & $1.0 \times 10^{0}$ & $1.1 \times 10^{0}$ \\
        & FF-PINN & $1.0 \times 10^{0}$ & $1.2 \times 10^{0}$ & $1.1 \times 10^{0}$ \\
        & PI-KAN & $1.0 \times 10^{0}$ & --- & --- \\
        & FB-PINN & $1.0 \times 10^{0}$ & --- & --- \\ \hline
        \multirow{6}{*}{\rotatebox[origin=c]{90}{FP32}} & PI-SUPN (ours) & $\mathbf{1.1 \times 10^{-4}}$ & $\mathbf{1.7 \times 10^{-3}}$ & $\mathbf{6.9 \times 10^{-1}}$ \\
        & PINN & $1.0 \times 10^{0}$ & $8.0 \times 10^{-1}$ & $1.1 \times 10^{0}$ \\
        & SIREN & $1.0 \times 10^{0}$ & $1.0 \times 10^{0}$ & $1.2 \times 10^{0}$ \\
        & FF-PINN & $1.0 \times 10^{0}$ & $9.9 \times 10^{-1}$ & $1.1 \times 10^{0}$ \\
        & PI-KAN & --- & --- & --- \\
        & FB-PINN & --- & --- & --- \\
        \bottomrule
    \end{tabular}
    \caption{Median relative $L^2$ error over 5 initializations for mode 1 of the tuning fork (\Cref{sec:tuning-fork}) at the best-performing architecture.}
    \label{tbl:tuning-fork-mode-1}
\end{table}

\begin{table}[h!]
    \renewcommand{\arraystretch}{1.1}
    \centering
    \begin{tabular}{l|c||c|c||c|c||c|c}
        \toprule
        & \multirow{2}{*}{Method} & \multicolumn{2}{c||}{Adam + SS-BFGS} & \multicolumn{2}{c||}{Adam + SOAP} & \multicolumn{2}{c}{Adam only} \\ \cline{3-8}
        & & Mode 2 & Mode 3 & Mode 2 & Mode 3 & Mode 2 & Mode 3 \\ \hline
        \multirow{2}{*}{\rotatebox[origin=c]{90}{FP64}} & PI-SUPN (ours) & $\mathbf{1.1 \times 10^{-6}}$ & $\mathbf{1.4 \times 10^{-5}}$ & $1.3 \times 10^{0}$ & $1.0 \times 10^{0}$ & $5.6 \times 10^{0}$ & $2.7 \times 10^{1}$ \\
        & PINN & $1.0 \times 10^{0}$ & $1.1 \times 10^{0}$ & $1.0 \times 10^{0}$ & $1.0 \times 10^{0}$ & $1.0 \times 10^{0}$ & $1.0 \times 10^{0}$ \\ \hline
        \multirow{2}{*}{\rotatebox[origin=c]{90}{FP32}} & PI-SUPN (ours) & $1.4 \times 10^{0}$ & $2.0 \times 10^{0}$ & $1.3 \times 10^{0}$ & $1.1 \times 10^{0}$ & $6.4 \times 10^{0}$ & $3.2 \times 10^{1}$ \\
        & PINN & $1.0 \times 10^{0}$ & $1.1 \times 10^{0}$ & $1.0 \times 10^{0}$ & $1.0 \times 10^{0}$ & $1.0 \times 10^{0}$ & $1.0 \times 10^{0}$ \\
        \bottomrule
    \end{tabular}
    \caption{Median relative $L^2$ error for modes 2 and 3 for the tuning fork (\Cref{sec:tuning-fork}) at the best-performing architecture. Based on the performance in \Cref{tbl:tuning-fork-mode-1}, only PINNs and PI-SUPNs are shown. Bold indicates the lowest error only when the relative error is below $10^{-1}$.}
    \label{tbl:tuning-fork-modes-2-and-3}
\end{table}

\subsection{2D nonlinear Swift-Hohenberg} \label{sec:sh}

\begin{figure}
    \centering
    \includegraphics[width=0.9\textwidth]{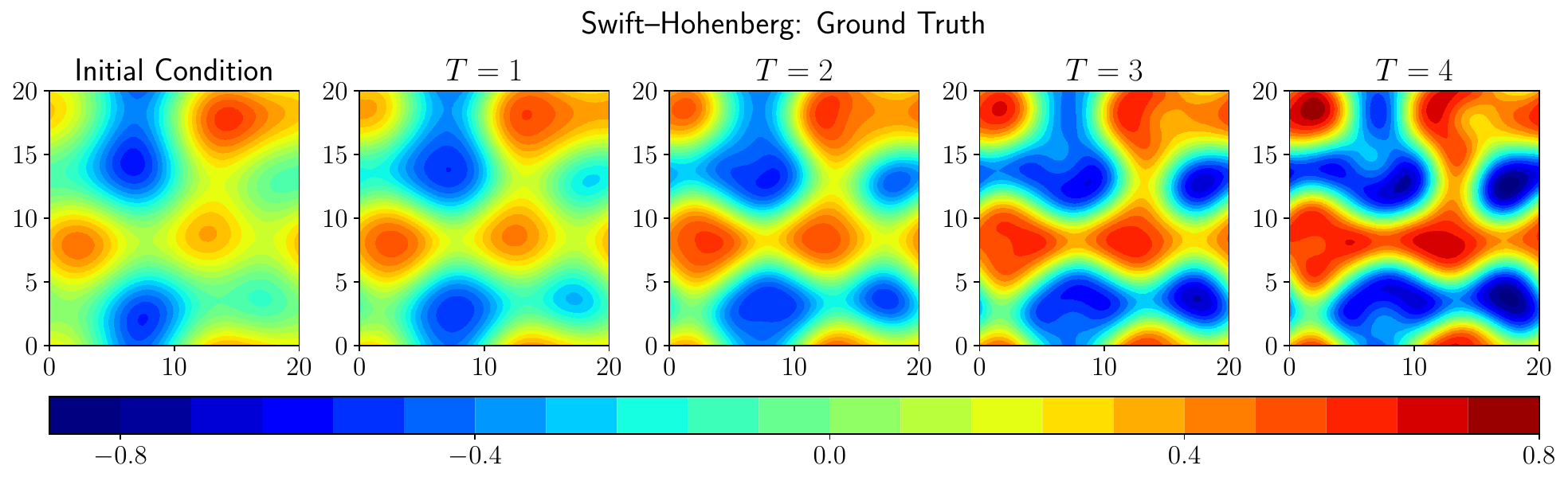}
    \caption{Initial condition and final-time reference solution for Swift--Hohenberg (\Cref{sec:sh}).}
    \label{fig:sh_ref}
\end{figure}

Consider the following fourth-order PDE problem:
\begin{align}
&\frac{\partial u}{\partial t}
= (\alpha-1)u
- \left(
2\frac{\partial^2 u}{\partial x^2}
+ 2\frac{\partial^2 u}{\partial y^2}
+ \frac{\partial^4 u}{\partial x^4}
+ 2\frac{\partial^4 u}{\partial x^2 \partial y^2}
+ \frac{\partial^4 u}{\partial y^4}
\right)
+ \beta u^2 - u^3, \\
& (x,y,t) \in [0,20] \times [0,20] \times (0,T], \nonumber \\
& u(x,y,0) = u_0(x,y), \nonumber \\
& \frac{\partial^k u}{\partial x^k}(0,y,t)
= \frac{\partial^k u}{\partial x^k}(20,y,t),
\qquad
\frac{\partial^k u}{\partial y^k}(x,0,t)
= \frac{\partial^k u}{\partial y^k}(x,20,t),
\quad k=0,\ldots,3.
\end{align}
where $\alpha = 0.8$ and $\beta = 0$. We consider time-marching~\cite{penwarden2023unified} to $T=4$ with a time step of $\Delta t =1$ using causality-enhanced discrete-time PINNs based on forward Euler~\cite{li2024causality}. Specifically, we fit a sequence of four physics-informed models $\hat{u}_k(x, y, t)$ on $t \in (0, 1]$ with the initial conditions
$$
\hat{u}_k(x, y, 0) = \begin{cases}
    u_0(x, y), \qquad & k = 0, \\
    \hat{u}_{k-1}(x, y, 1),  & k = 1, 2, 3
\end{cases} \ .
$$
We take $u_0(x, y)$ as a sample of a periodic Gaussian random field with length-scale 10, i.e. half the domain length, shown in \Cref{fig:sh_ref}. To generate high-fidelity test data (only for measuring approximation error), we use the implementation of Swift--Hohenberg in Chebfun~\cite{driscoll2014chebfun} based on exponential time-differencing~\cite{cox2002exponential}. Generating snapshot data at $T \in \{1, 2, 3, 4\}$ on a $129 \times 129$ spatial mesh with Chebfun takes 0.1 seconds.

\begin{figure}
    \centering
    \includegraphics[width=0.8\textwidth]{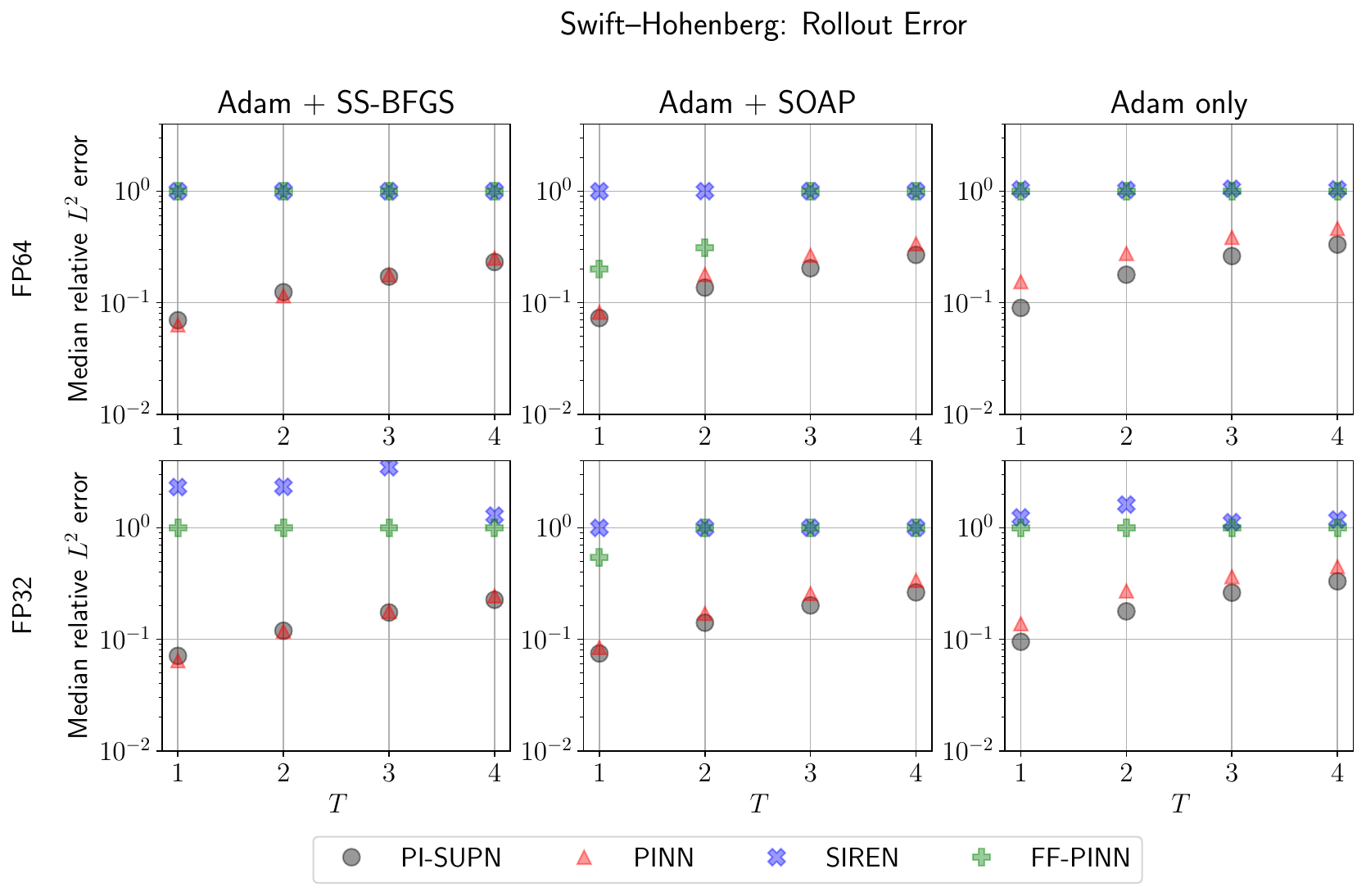}
    \caption{Approximation error at various final times $T$ for Swift--Hohenberg (\Cref{sec:sh}) using the best-performing architecture from different combinations of precision and optimizer.}
    \label{fig:sh_rollout_error}
\end{figure}

We choose the architecture for the eventual rollout based on an ablation study for a single time step to $T=1$, discussed in \Cref{app:sh}. Due to the runtime overhead for PI-KANs and FB-PINNs shown in previous experiments, we omit them here. \Cref{fig:sh_rollout_error} shows that PI-SUPNs have a slight edge over PINNs at $T=4$ regardless of optimizer or precision, although PINNs obtain slightly lower error at $T=1$ using Adam + SS-BFGS. In all cases, however, PINNs take approximately 50\% longer to train than PI-SUPNs (see \Cref{app:sh}). Additionally, Fourier-feature PINNs exhibit error that is optimizer/precision-dependent as well as unstable in time, while SIRENs roundly struggle. We attribute this phenomenon to the time dimension being passed through $\sin$ or $\cos$, whereas the dynamics in \Cref{fig:sh_ref} are not periodic in time.

%% file: summary.tex
\section{Conclusions}

In this work, we investigated Physics-Informed Shallow Universal Polynomial Networks (PI-SUPNs) as an alternative to conventional deep architectures for solving challenging PDEs. Our results demonstrate that architectural shallowness can substantially improve trainability without sacrificing approximation capacity, while reducing the computational overhead associated with repeated differentiation. Although curvature-aware optimization and FP64 precision improve convergence, they cannot always overcome optimization difficulties arising from the underlying network architecture. Conversely, our high-order PDE experiments show that architectural improvements alone are insufficient, with successful training requiring an appropriate combination of architecture, numerical precision, and optimizer. These findings suggest that optimization stiffness in physics-informed learning should be addressed through architectural design {\it in conjunction with} more sophisticated optimization methods rather than relying on them exclusively. Future work will investigate the relationship between network architecture and loss conditioning, as well as extending SUPNs to physics-informed neural operators.

\section*{Acknowledgments}
The research described in this paper was conducted under the auspices of Sandia National Laboratories (SNL, Contract No. DE-NA0003525), Lawrence Livermore National Laboratory (LLNL, Contract No. DE-AC52-07NA27344), and Pacific Northwest National Laboratory (PNNL, Contract No. DE-AC05--76RL01830) as part of the DOE Interlaboratory (IL) Laboratory Directed Research and Development (LDRD) Pilot Program. The LLNL LDRD project number is 25-ERD-052, and the release number is LLNL-JRNL-2026550.
Any subjective views or opinions that might be expressed in the paper do not necessarily represent the views of the U.S. Department of Energy or the United States Government. The publisher acknowledges that the U.S. Government retains a non-exclusive, paid-up, irrevocable, worldwide license to publish or reproduce the published form of this written work or allow others to do so, for U.S. Government purposes. The DOE will provide public access to results of federally sponsored research in accordance with the DOE Public Access Plan.

%% file: appendix.tex
\section{Efficient Polynomial Calculations}
\label{app:efficent_polys}
Three-term recurrence relations for polynomials are highly efficient for evaluating a polynomial basis with $O(N)$, rather than $O(N^2)$, work. For Chebyshev polynomials, the recurrence relation is
\begin{align*}
    T_0(x) &= 1 \\
    T_1(x) &= x \\
    T_{n+1}(x) &= 2x \, T_n(x) - T_{n-1}(x), \qquad n \geq 1
\end{align*}
where $x \in [-1, 1]$. 

\subsection{Derivatives}

Propagating automatic differentiation through each Chebyshev polynomial is needlessly (and prohibitively) expensive. We can precompute Chebyshev derivatives at the collocation points, Neumann BCs, or Robin BCs. Even this precomputation can be done efficiently with the three-term recurrence. Taking the first derivative gives the base case
\begin{align*}
    T_0'(x) &= 0 \\
    T_1'(x) &= 1 \\
    T_{n+1}'(x) &= 2x \, T_n'(x) + 2T_n(x) - T_{n-1}'(x) \, .
\end{align*}
Taking further derivatives gives the recursion
\begin{equation}
        T_{n+1}^{(r)}(x) = 2x \, T_n^{(r)}(x) + 2r \, T_n^{(r-1)}(x) - T_{n-1}^{(r)}(x), \qquad r \geq 1, n \geq 1
\end{equation}

\subsection{Avoiding Redundant Univariate Evaluations}

In the multidimensional case, the polynomial degrees are given by a downward-closed multi-index set $\Lambda \subset \N_0^d$. Many of the multi-indices will share univariate degrees. For example, the total-degree space of order 5 contains both $T_2(x)T_3(y)$ and $T_2(x) T_2(y)$, and we do not need to compute $T_2(x)$ more than once.

We avoid redundant calculations by calculating all needed polynomial degrees for each dimension at the front-end and then multiplying them together as needed. Define
$$
\vec{\lambda}_k \coloneqq \max_{\vec{i} \in \Lambda} \vec{i}_k
$$
For each dimension $1 \leq k \leq d$, we compute and store all $T_j(x_k)$ with $j \leq \lambda_k$. Then we build our multivariate basis matrix by iterating over $\Lambda$ and multiplying precomputed univariate polynomials.

\section{Supporting Approximation and Convergence Theory}
\label{app:additional_proofs}

We give the proof of the quasi-local approximation theorem stated in
\Cref{thm:spline-interpolant-supn} and derive an accompanying spline-based
approximation bound. Throughout, $\Omega\subset\mathbb{R}$ is a compact
interval, $\mathbb{P}_M$ denotes the polynomials of degree at most $M$,
and
\begin{align}
\epsilon_M(g;W^{r,p}(\Omega))
&:=\inf_{q\in\mathbb{P}_M}\|g-q\|_{W^{r,p}(\Omega)}.
\label{eq:app-best-poly-error}
\end{align}
We use the equivalent Sobolev norm
$\|g\|_{W^{r,p}(\Omega)}=\max_{0\leq j\leq r}\|g^{(j)}\|_{L^p(\Omega)}$;
this is equivalent to the usual sum norm for fixed $r$.
A B-spline of degree $k$ belongs to $W^{k,p}(\Omega)$ for
$1\leq p\leq\infty$; its conventional spline order is $k+1$.

\subsection{Proof of quasi-local approximation}
\label{app:quasi-local-proof}

\begin{lemma}[Bounded tails]
\label{thm:supn-bounded-tails}
Let $B\in W^{r,p}(\Omega)$ vanish outside a closed set
$\Omega_B\subset\Omega$. Then, for every $v\in W^{r,p}(\Omega)$,
\begin{align}
\|v\|_{W^{r,p}(\Omega\setminus\Omega_B)}
&\leq\|B-v\|_{W^{r,p}(\Omega)}.
\label{eq:app-bounded-tails}
\end{align}
\end{lemma}
\begin{proof}
The weak derivatives of $B$ of orders $0,\ldots,r$ vanish almost
everywhere on $\Omega\setminus\Omega_B$. Hence, for every such order $j$,
\begin{align*}
\|v^{(j)}\|_{L^p(\Omega\setminus\Omega_B)}
&=\|(B-v)^{(j)}\|_{L^p(\Omega\setminus\Omega_B)}\\
&\leq\|(B-v)^{(j)}\|_{L^p(\Omega)}.
\end{align*}
Taking the maximum over $j$ proves the claim, including $p=\infty$.
\end{proof}

\begin{proof}[Proof of \Cref{thm:spline-interpolant-supn}]
Write $s_{N,k}=\sum_{n=0}^{N-1}w_nB_{n,k}$, where each
$B_{n,k}$ is a degree-$k$ B-spline, and fix
$0\leq r\leq k$, $1\leq p\leq\infty$, and $\delta>0$.
Let $q^*_{n,M}\in\mathbb{P}_M$ attain the best approximation error
\begin{align*}
\epsilon_{n,M}
:=\|B_{n,k}-q^*_{n,M}\|_{W^{r,p}(\Omega)}
=\epsilon_M(B_{n,k};W^{r,p}(\Omega)).
\end{align*}
Assume, as in the main-text theorem, that each nonzero spline has a
support with nonempty open complement in $\Omega$. Then
$\epsilon_{n,M}>0$: if the error vanished, a polynomial would
agree with the spline almost everywhere, and would therefore vanish
on an open interval while being nonzero elsewhere. The identically
zero spline can be omitted from the expansion.

The polynomial-to-SUPN approximation result of
\cite[Theorem~3.3]{morrow2026supn} implies that the polynomial
$q^*_{n,M}$ can be approximated arbitrarily closely in
$W^{r,\infty}(\Omega)$ by a single-neuron, degree-$M$ SUPN.
Choose such a neuron $\hat f_{1,M}^{n,k}$ so that
\begin{align}
\|q^*_{n,M}-\hat f_{1,M}^{n,k}\|_{W^{r,p}(\Omega)}
&\leq\delta\epsilon_{n,M}.
\label{eq:app-polynomial-neuron-proximity}
\end{align}
Here the $W^{r,p}$ estimate follows from the
$W^{r,\infty}$ approximation on the bounded interval, with the
necessary measure factor absorbed in the arbitrarily chosen tolerance.
Consequently, the triangle inequality gives
\begin{align}
\|B_{n,k}-\hat f_{1,M}^{n,k}\|_{W^{r,p}(\Omega)}
&\leq(1+\delta)\epsilon_{n,M}.
\label{eq:app-individual-spline-approximation}
\end{align}
Applying \Cref{thm:supn-bounded-tails} with
$\Omega_B=\Omega_{n,k}:=\operatorname{supp}(B_{n,k})$ yields
\begin{align}
\|\hat f_{1,M}^{n,k}\|_{W^{r,p}(\Omega\setminus\Omega_{n,k})}
&\leq(1+\delta)\epsilon_{n,M}.
\label{eq:app-individual-tail}
\end{align}
Finally, define $\hat f_{N,M}:=\sum_{n=0}^{N-1}w_n\hat f_{1,M}^{n,k}$.
By the triangle inequality,
\begin{align}
\|s_{N,k}-\hat f_{N,M}\|_{W^{r,p}(\Omega)}
&\leq\sum_{n=0}^{N-1}|w_n|
\|B_{n,k}-\hat f_{1,M}^{n,k}\|_{W^{r,p}(\Omega)}\nonumber\\
&\leq(1+\delta)\sum_{n=0}^{N-1}|w_n|
\epsilon_M(B_{n,k};W^{r,p}(\Omega)).
\label{eq:app-quasi-local-total}
\end{align}
These are the asserted estimates.
\end{proof}

\begin{rmk}[Zero polynomial-approximation error]
\label{rmk:zero-approx-error}
The positive-error condition is needed only to obtain the stated
multiplicative estimates for a finite SUPN. If
$\epsilon_M(B_{n,k};W^{r,p}(\Omega))=0$, the polynomial-to-SUPN
result instead gives, for every $\eta>0$, a finite neuron with
$\|B_{n,k}-\hat f_{1,M}^{n,k}\|_{W^{r,p}(\Omega)}<\eta$.
The quasi-local and weighted-sum bounds then hold with an additive
arbitrarily small error. Sending $\eta\to0$ yields the corresponding
bound for the \emph{infimum} over SUPN parameters, but does not
assert attainment by a finite parameter choice. This distinction does
not affect the strict-support B-splines used above.
\end{rmk}

\subsection{Spline-based SUPN approximation convergence}
\label{app:supn-spline-convergence}

The following bound makes explicit the additional stability and
approximation assumptions required to connect polynomial degree and
network width. It is an existence result for the SUPN approximation
class, not a guarantee about the outcome of physics-informed training.

\begin{theorem}[Spline-based SUPN approximation bound]
\label{thm:supn-spline-convergence}
Let $k\geq1$ and let $r,s$ be integers satisfying
$0\leq r<s\leq k+1$ and $r<k$. Let $1\leq p\leq\infty$ and
$f\in W^{s,p}(\Omega)$. Suppose that, for each sufficiently large $N$,
there exists a degree-$k$ spline quasi-interpolant
\begin{align}
s_{N,k}=\sum_{n=0}^{N-1}w_nB_{n,k}
\end{align}
on a quasi-uniform knot sequence of mesh size $h\asymp N^{-1}$,
with B-spline supports $\Omega_{n,k}$ that have nonempty open complements in $\Omega$, such that
\begin{align}
\|f-s_{N,k}\|_{W^{r,p}(\Omega)}
&\leq C_sN^{-(s-r)}\|f\|_{W^{s,p}(\Omega)},
\label{eq:app-spline-error}\\
|w_n|&\leq C_wN^{1/p}
\|f\|_{L^p(\Omega_{n,k})},
\label{eq:app-spline-stability}
\end{align}
where $N^{1/\infty}:=1$ and the constants are independent of $N$.
Assume also the standard polynomial approximation estimate
\begin{align}
\epsilon_M(g;W^{r,p}(\Omega))
&\leq C_JM^{-(k-r)}\|g\|_{W^{k,p}(\Omega)}
\label{eq:app-polynomial-jackson}
\end{align}
for the degree-$k$ spline functions considered, with $C_J$
independent of $N$ and $M$. Then, for each $\delta>0$ and $M\geq1$,
there exists an $N$-neuron, degree-$M$ SUPN for which
\begin{align}
\|f-\hat f_{N,M}\|_{W^{r,p}(\Omega)}
&\leq\alpha N^{-(s-r)}\|f\|_{W^{s,p}(\Omega)}\nonumber\\
&\quad+\beta(1+\delta)N^{k+1-1/p}M^{-(k-r)}
\|f\|_{L^p(\Omega)},
\label{equ:supn-spline-convergence}
\end{align}
where $\alpha,\beta>0$ are independent of $N$ and $M$.
\end{theorem}

\begin{proof}
The triangle inequality and \Cref{eq:app-quasi-local-total} give
\begin{align}
\|f-\hat f_{N,M}\|_{W^{r,p}(\Omega)}
&\leq\|f-s_{N,k}\|_{W^{r,p}(\Omega)}\nonumber\\
&\quad+(1+\delta)\sum_{n=0}^{N-1}|w_n|
\epsilon_M(B_{n,k};W^{r,p}(\Omega)).
\label{eq:app-error-decomposition}
\end{align}
The first term is bounded by \Cref{eq:app-spline-error}.
For the second, a degree-$k$ B-spline on a quasi-uniform mesh satisfies
\begin{align}
\|B_{n,k}^{(j)}\|_{L^p(\Omega)}
&\leq C_BN^{j-1/p},\qquad 0\leq j\leq k,
\label{eq:app-bspline-scaling}
\end{align}
where the constant may depend on $k$, $p$, and the mesh
quasi-uniformity but not on $n$ or $N$. Thus, since $N\geq1$,
\begin{align}
\|B_{n,k}\|_{W^{k,p}(\Omega)}
&\leq C_BN^{k-1/p},
\end{align}
and \Cref{eq:app-polynomial-jackson} gives
\begin{align}
\epsilon_M(B_{n,k};W^{r,p}(\Omega))
&\leq C_JC_BM^{-(k-r)}N^{k-1/p}.
\label{eq:app-bspline-poly-rate}
\end{align}
There are at most $k+1$ overlapping degree-$k$ B-spline supports
at almost every point, and the coefficient bound
\Cref{eq:app-spline-stability} implies
\begin{align}
\sum_{n=0}^{N-1}|w_n|
&\leq C_wN^{1/p}\sum_{n=0}^{N-1}
\|f\|_{L^p(\Omega_{n,k})}\nonumber\\
&\leq C N\|f\|_{L^p(\Omega)}.
\label{eq:app-coefficient-sum}
\end{align}
For $1\leq p<\infty$, the second inequality follows from
H\"older's inequality and bounded support overlap; for $p=\infty$
it follows by summing the uniform coefficient bounds.
Inserting \Cref{eq:app-bspline-poly-rate,eq:app-coefficient-sum}
into \Cref{eq:app-error-decomposition} proves
\Cref{equ:supn-spline-convergence} with constants independent
of $N$ and $M$.
\end{proof}

\begin{rmk}[Spline approximation assumptions]
\label{rmk:spline-assumptions}
\Cref{thm:supn-spline-convergence} is conditional on the stated
Sobolev approximation and local coefficient-stability estimates for
the selected degree-$k$ spline quasi-interpolant, and on the
simultaneous Sobolev polynomial approximation estimate
\eqref{eq:app-polynomial-jackson}. The constants in these estimates
must be uniform in $N$ and $M$; they are not claimed to hold for an
arbitrary spline representation. The restriction $r<k$ ensures a
positive algebraic rate $M^{-(k-r)}$.
\end{rmk}

\begin{corollary}[Joint refinement of width and degree]
\label{cor:supn-convergence}
Under the assumptions of \Cref{thm:supn-spline-convergence},
let $M\geq AN^\gamma$ for some $A>0$. The approximation-error
bound tends to zero as $N\to\infty$ whenever
\begin{align}
\gamma&>\frac{k+1-1/p}{k-r}.
\label{eq:app-convergence-condition}
\end{align}
The two terms in the error bound have the same asymptotic order
when
\begin{align}
\gamma&=\frac{k+1-1/p+s-r}{k-r}.
\label{eq:app-balanced-refinement}
\end{align}
\end{corollary}
\begin{proof}
Substituting $M\geq AN^\gamma$ into
\Cref{equ:supn-spline-convergence} bounds its right-hand side
by a constant multiple of
\begin{align*}
N^{-(s-r)}+N^{k+1-1/p-\gamma(k-r)}.
\end{align*}
Both exponents are negative under
\Cref{eq:app-convergence-condition}; equating the exponents
gives \Cref{eq:app-balanced-refinement}.
\end{proof}

\noindent\textit{Scope.}
These results provide an approximation-existence bound for a
one-dimensional SUPN ansatz. They do not prove PDE-residual
convergence, convergence of a training algorithm, or improved
conditioning of a physics-informed loss landscape.

\FloatBarrier

\section{Additional Experiments}

\subsection{Additional Helmholtz Results} \label{app:helmholtz}

\begin{figure}[!htbp]
    \centering
    \includegraphics[width=0.68\textwidth]{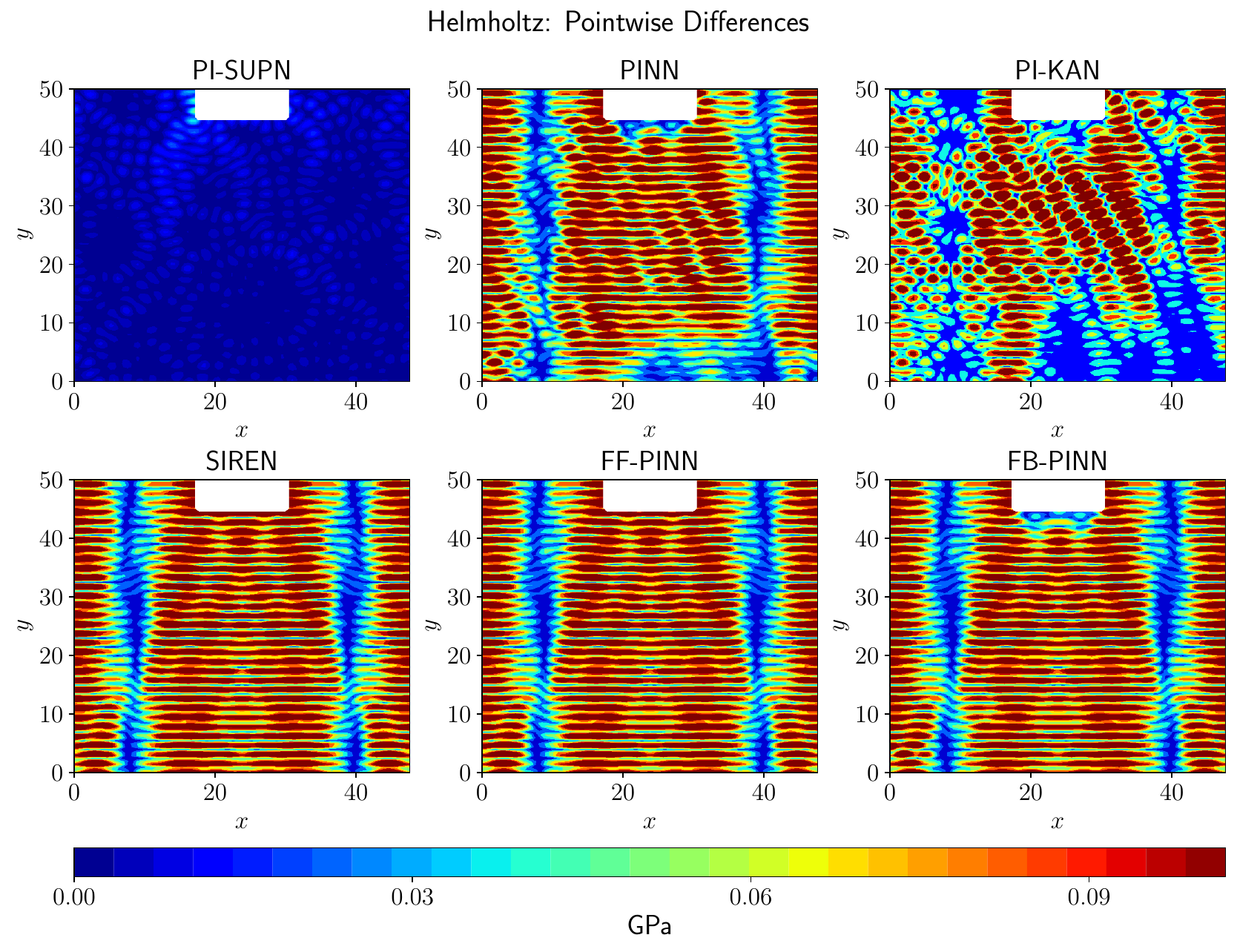}
    \caption{Pointwise absolute error for the Helmholtz problem (\Cref{sec:helmholtz}) using the median-selected models from \Cref{fig:helmholtz-convergence}.}
    \label{fig:helmholtz_pointwise_error}
\end{figure}

\begin{figure}[!htbp]
    \centering
    \includegraphics[width=0.68\textwidth]{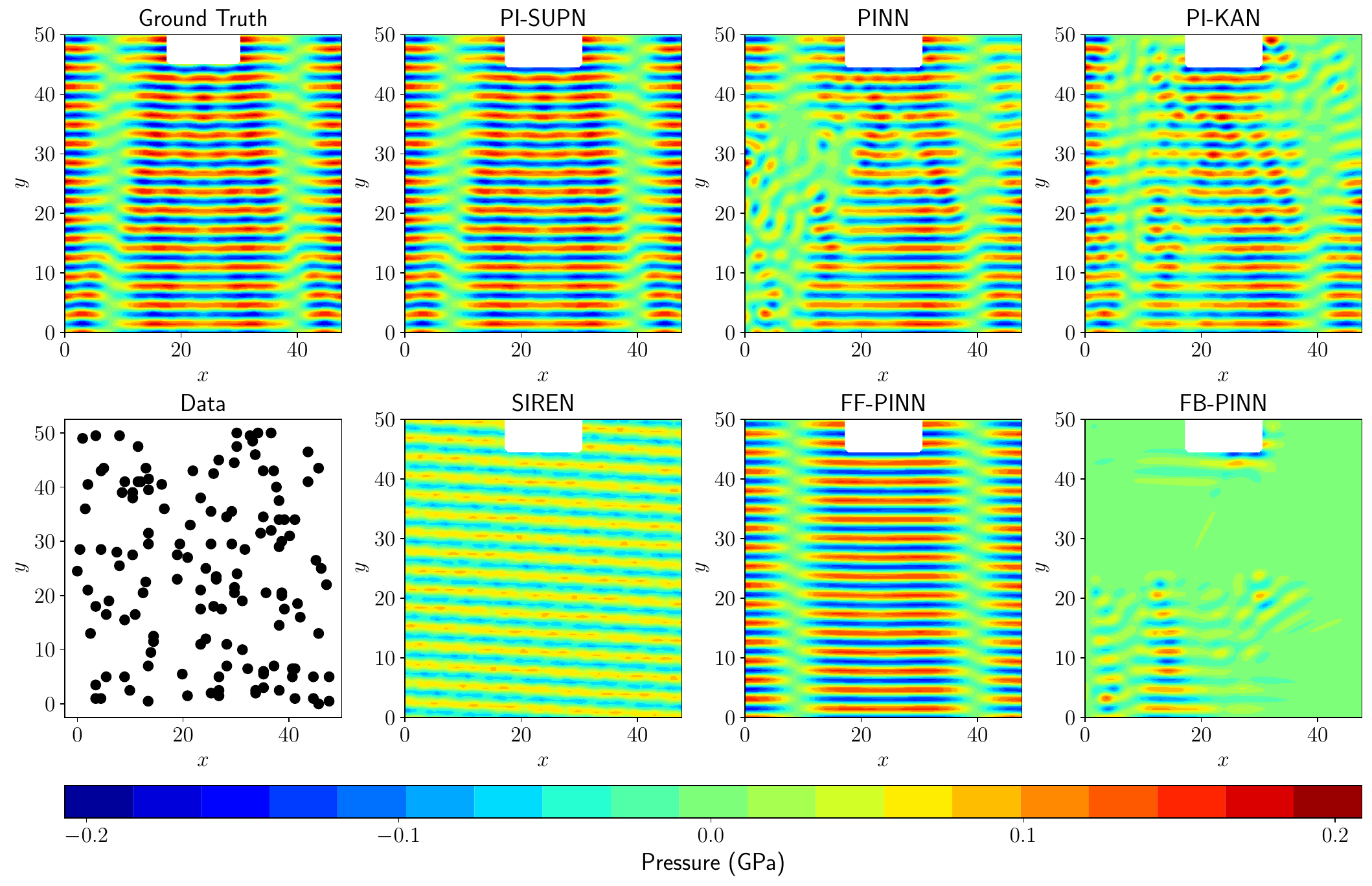}
    \caption{Best-performing Helmholtz predictions across optimizers, precisions, architectures, and network initializations.}
    \label{fig:helmholtz_4.75E5_min}
    \par\medskip
    \includegraphics[width=0.68\textwidth]{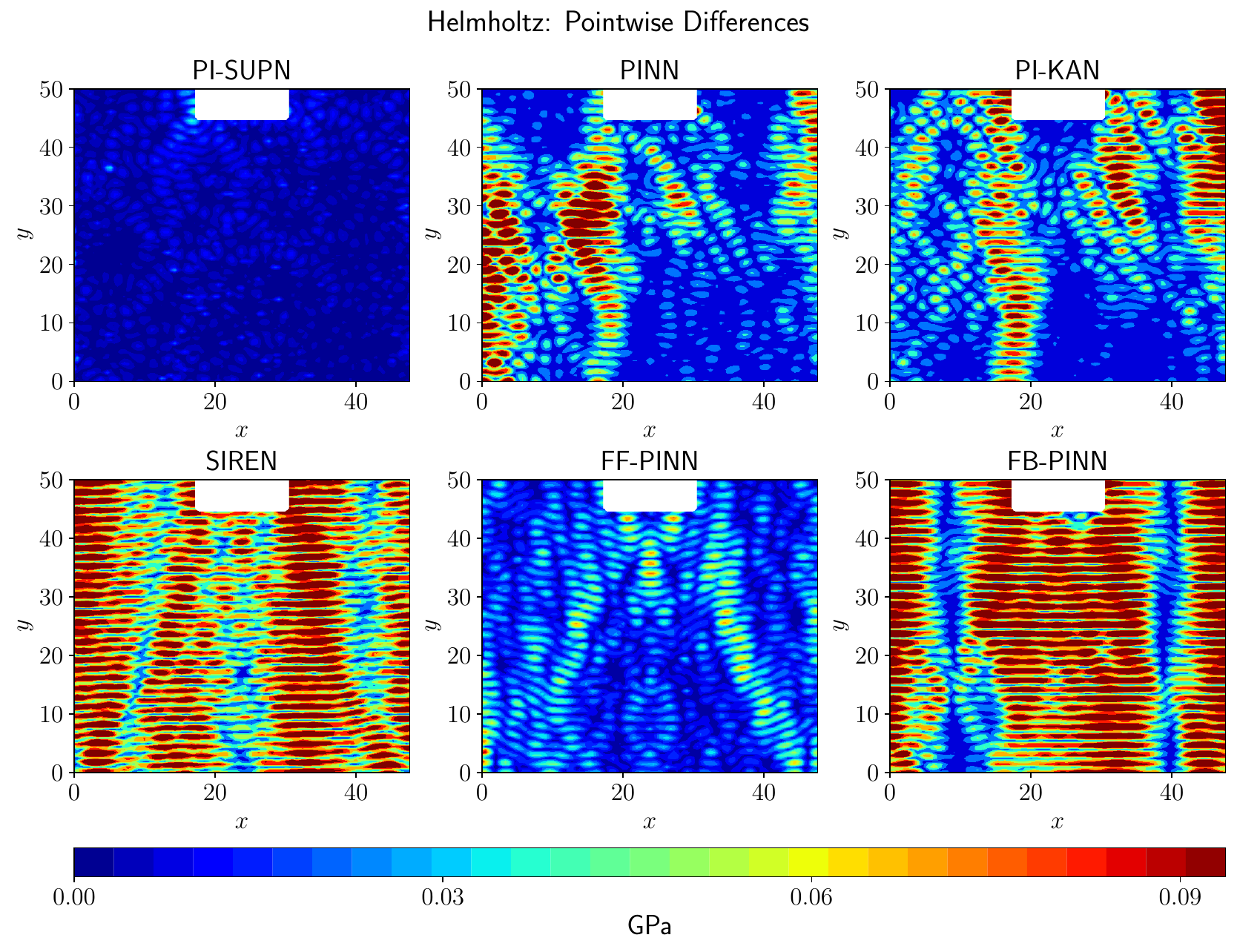}
    \caption{Pointwise absolute errors for the best-performing Helmholtz predictions in \Cref{fig:helmholtz_4.75E5_min}.}
    \label{fig:helmholtz_pointwise_error_min}
\end{figure}

COMSOL solves the boundary-value problem with second-order elements on a mesh with $3.3 \times 10^5$ nodes, in 10 seconds on a 14-core Intel Core i9-7940X CPU. \Cref{fig:helmholtz_pointwise_error} reports pointwise absolute errors for the median-selected Helmholtz models. \Cref{fig:helmholtz_4.75E5_min,fig:helmholtz_pointwise_error_min} show the predictions and corresponding pointwise errors for the best-performing models across initializations and training configurations.

\FloatBarrier
\subsection{Additional Allen--Cahn Results}
\label{app:allen-cahn-additional}
The following tables provide the full optimizer/precision accuracy ablations for the hard and easy Allen--Cahn benchmarks and the per-derivative errors for the hard case.

\begin{table}[!htbp]
    \centering
    \begin{tabular}{c|c||c|c|c|c}
    \toprule 
    & Method & Params & Adam + SS-BFGS & Adam + SOAP & Adam only \\ \hline
    \multirow{6}{*}{\rotatebox[origin=c]{90}{FP64}} & PI-SUPN (ours) & \textbf{2200} & $\mathbf{2.8 \times 10^{-2}}$ & $\mathbf{4.1 \times 10^{-1}}$ & $\mathbf{2.9 \times 10^{-1}}$ \\
     & PINN & 2210 & $1.1 \times 10^{-1}$ & $7.9 \times 10^{-1}$ & $9.3 \times 10^{-1}$ \\
     & SIREN & 2496 & $1.6 \times 10^{-1}$ & $6.4 \times 10^{-1}$ & $6.9 \times 10^{-1}$ \\
     & FF-PINN & 2208 & $6.1 \times 10^{-2}$ & $8.0 \times 10^{-1}$ & $9.4 \times 10^{-1}$ \\
     & PI-KAN & 3448 & $2.6 \times 10^{-1}$ & --- & --- \\
     & FB-PINN & 3968 & $7.3 \times 10^{-1}$ & --- & --- \\
    \hline
    \multirow{6}{*}{\rotatebox[origin=c]{90}{FP32}} & PI-SUPN (ours) & \textbf{2200} & $\mathbf{4.2 \times 10^{-2}}$ & $\mathbf{3.9 \times 10^{-1}}$ & $\mathbf{2.9 \times 10^{-1}}$ \\
     & PINN & 2210 & $5.7 \times 10^{-1}$ & $7.8 \times 10^{-1}$ & $9.4 \times 10^{-1}$ \\
     & SIREN & 2496 & $4.0 \times 10^{-1}$ & $7.5 \times 10^{-1}$ & $8.0 \times 10^{-1}$ \\
     & FF-PINN & 2208 & $8.7 \times 10^{-1}$ & $8.8 \times 10^{-1}$ & $8.7 \times 10^{-1}$ \\
     & PI-KAN & 3448 & $3.7 \times 10^{-1}$ & --- & --- \\
     & FB-PINN & 3968 & $1.0 \times 10^{0}$ & --- & --- \\
     \bottomrule 
    \end{tabular}
    \caption{Median relative $L^2$ error, over 5 initializations, of best-performing model on the hard Allen--Cahn problem (\Cref{sec:allen-cahn-hard}). Due to runtime, we only consider PI-KANs and FB-PINNs for the highest-accuracy optimizer.}
    \label{tbl:ac_hard_ablation}
\end{table}

\begin{table}[!htbp]
    \centering
    \begin{tabular}{c||c|c|c|c|c}
        \toprule
        Method & $u$ & $u_x$ & $u_y$ & $u_{xx}$ & $u_{yy}$ \\ \hline
        PI-SUPN (ours) & $\bm{9.3 \times 10^{-3}}$ & $\bm{6.9 \times 10^{-3}}$ & $\bm{1.2 \times 10^{-2}}$ & $\bm{7.7 \times 10^{-3}}$ & $\bm{1.4 \times 10^{-2}}$ \\
        PINN & $2.9 \times 10^{-1}$ & $6.2 \times 10^{-2}$ & $2.0 \times 10^{-1}$ & $2.7 \times 10^{-2}$ & $7.0 \times 10^{-2}$ \\
        SIREN & ${2.7 \times 10^{-2}}$ & ${1.5 \times 10^{-2}}$ & ${2.5 \times 10^{-2}}$ & ${1.8 \times 10^{-2}}$ & ${2.8 \times 10^{-2}}$ \\
        FF-PINN & $5.6 \times 10^{-2}$ & $2.5 \times 10^{-2}$ & $4.7 \times 10^{-2}$ & $2.4 \times 10^{-2}$ & $4.3 \times 10^{-2}$ \\
        PI-KAN & $3.7 \times 10^{-1}$ & $8.0 \times 10^{-2}$ & $1.6 \times 10^{-1}$ & $9.7 \times 10^{-2}$ & $1.4 \times 10^{-1}$ \\
        FB-PINN & $1.0 \times 10^{0}$ & $4.9 \times 10^{-1}$ & $7.5 \times 10^{-1}$ & $4.4 \times 10^{-1}$ & $7.9 \times 10^{-1}$ \\
        \bottomrule 
    \end{tabular}
    \caption{Median $L^2$ errors for each spatial derivative on the hard Allen--Cahn problem (\Cref{sec:allen-cahn-hard}) using Adam + SS-BFGS with FP32.}
    \label{tbl:ac_hard_derivatives}
\end{table}

\begin{table}[!htbp]
    \centering
    \begin{tabular}{c|c||c|c|c|c}
    \toprule 
    & Method & Params & Adam + SS-BFGS & Adam + SOAP & Adam only \\ \hline
    \multirow{6}{*}{\rotatebox[origin=c]{90}{FP64}} & PI-SUPN (ours) & $\bm{3000}$ & $1.1 \times 10^{-5}$ & $\bm{6.9 \times 10^{-4}}$ & $\bm{3.0 \times 10^{-3}}$ \\
     & PINN & 3296 & ${8.0 \times 10^{-6}}$ & $6.5 \times 10^{-3}$ & $5.0 \times 10^{-2}$ \\
     & SIREN & 3296 & $9.8 \times 10^{-6}$ & ${1.8 \times 10^{-3}}$ & ${9.6 \times 10^{-3}}$ \\
     & FF-PINN & ${3232}$ & $\bm{4.0 \times 10^{-6}}$ & $5.2 \times 10^{-3}$ & $1.4 \times 10^{-2}$ \\
     & PI-KAN & 3448 & $1.9 \times 10^{-3}$ & $9.2 \times 10^{-3}$ & $1.0 \times 10^{-2}$ \\
     & FB-PINN & 3968 & $3.1 \times 10^{-5}$ & $4.6 \times 10^{-3}$ & $1.4 \times 10^{-2}$ \\
    \hline
    \multirow{6}{*}{\rotatebox[origin=c]{90}{FP32}} & PI-SUPN (ours) & $\bm{3000}$ & $\bm{3.1 \times 10^{-5}}$ & $\bm{7.8 \times 10^{-4}}$ & $\bm{3.0 \times 10^{-3}}$ \\
     & PINN & 3296 & $5.2 \times 10^{-4}$ & $6.1 \times 10^{-3}$ & $4.5 \times 10^{-2}$ \\
     & SIREN & 3296 & ${3.4 \times 10^{-5}}$ & ${1.8 \times 10^{-3}}$ & ${5.9 \times 10^{-3}}$ \\
     & FF-PINN & ${3232}$ & $1.5 \times 10^{-4}$ & $5.8 \times 10^{-3}$ & $1.3 \times 10^{-2}$ \\
     & PI-KAN & 3448 & $1.8 \times 10^{-3}$ & $6.0 \times 10^{-3}$ & $7.3 \times 10^{-3}$ \\
     & FB-PINN & 3968 & $2.2 \times 10^{-3}$ & $5.1 \times 10^{-3}$ & $1.5 \times 10^{-2}$ \\
    \bottomrule 
    \end{tabular}
    \caption{Median relative $L^2$ error, over 5 initializations, of best-performing model on the easy Allen--Cahn problem (\Cref{sec:allen-cahn-easy}). In each case, SUPNs have the fewest trainable parameters.}
    \label{tbl:ac_easy_ablation}
\end{table}

\FloatBarrier
\subsection{Additional Euler--Bernoulli Beam Results}
\label{app:beam-additional}
Training times for the fundamental tuning-fork mode are reported below; optimizer and precision accuracy comparisons appear in \Cref{tbl:tuning-fork-mode-1}.

\begin{table}[!htbp]
    \centering
    \begin{tabular}{l|c||c|c|c|c}
        \toprule
        & Method & Num. params. & Adam + SS-BFGS & Adam + SOAP & Adam only \\ \hline
        \multirow{6}{*}{\rotatebox[origin=c]{90}{FP64}} & PI-SUPN (ours) & 1400 & $\bm{2.3 \times 10^{-2}}$ & $\bm{1.2 \times 10^{-2}}$ & $\bm{1.1 \times 10^{-2}}$ \\
        & PINN & 1424 & ${6.7 \times 10^{-2}}$ & $5.6 \times 10^{-2}$ & $5.2 \times 10^{-2}$ \\
        & SIREN & 1424 & $8.3 \times 10^{-2}$ & $5.5 \times 10^{-2}$ & $5.1 \times 10^{-2}$ \\
        & FF-PINN & 1392 & $8.7 \times 10^{-2}$ & ${5.5 \times 10^{-2}}$ & ${5.0 \times 10^{-2}}$ \\
        & PI-KAN & 1232 & $5.3 \times 10^{-1}$ & ---  & ---  \\
        & FB-PINN & 1024 & $3.1 \times 10^{0}$ & ---  & ---  \\ \hline
        \multirow{4}{*}{\rotatebox[origin=c]{90}{FP32}} & PI-SUPN (ours) & 1400 & $\bm{9.3 \times 10^{-3}}$ & $\bm{7.8 \times 10^{-3}}$ & $\bm{7.2 \times 10^{-3}}$ \\
        & PINN & 1424 & ${5.9 \times 10^{-2}}$ & ${5.3 \times 10^{-2}}$ & $5.1 \times 10^{-2}$ \\
        & SIREN & 1424 & $6.4 \times 10^{-2}$ & $5.5 \times 10^{-2}$ & $5.1 \times 10^{-2}$ \\
        & FF-PINN & 1392 & $6.5 \times 10^{-2}$ & $5.4 \times 10^{-2}$ & ${5.1 \times 10^{-2}}$ \\
        \bottomrule
    \end{tabular}
    \caption{Median training time per epoch (sec) for networks of comparable sizes on the first mode of the tuning fork (\Cref{sec:tuning-fork}).}
    \label{tbl:tuning-fork-runtime}
\end{table}

\FloatBarrier
\subsection{Additional Swift--Hohenberg Results} \label{app:sh}

\begin{figure}[!htbp]
    \centering
    \includegraphics[width=0.8\textwidth]{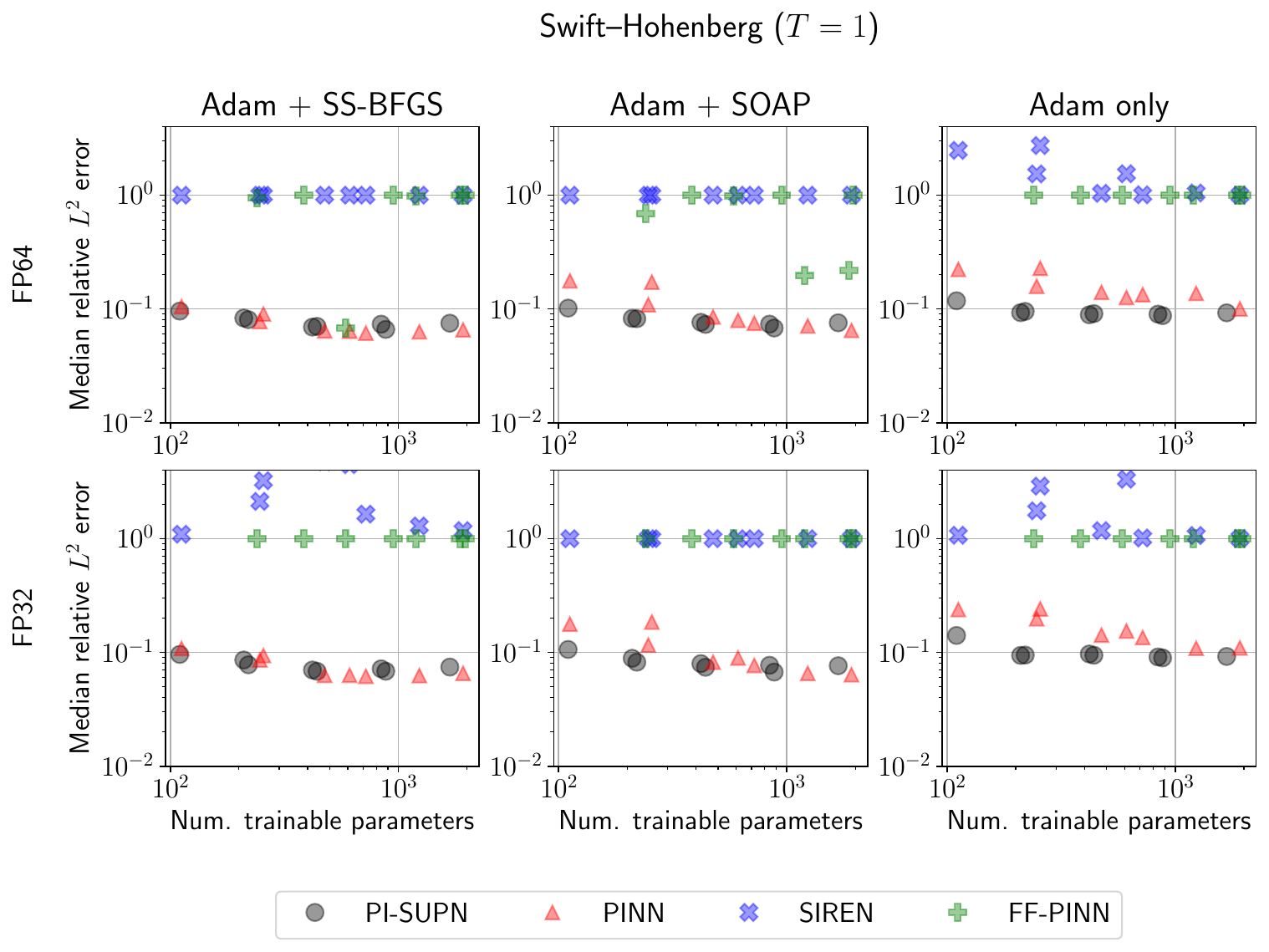}
    \caption{Ablation study for Swift--Hohenberg (\Cref{sec:sh}) at $t=1$.}
    \label{fig:sh_t_1}
\end{figure}

\begin{figure}[!htbp]
    \centering
    \includegraphics[width=0.9\textwidth]{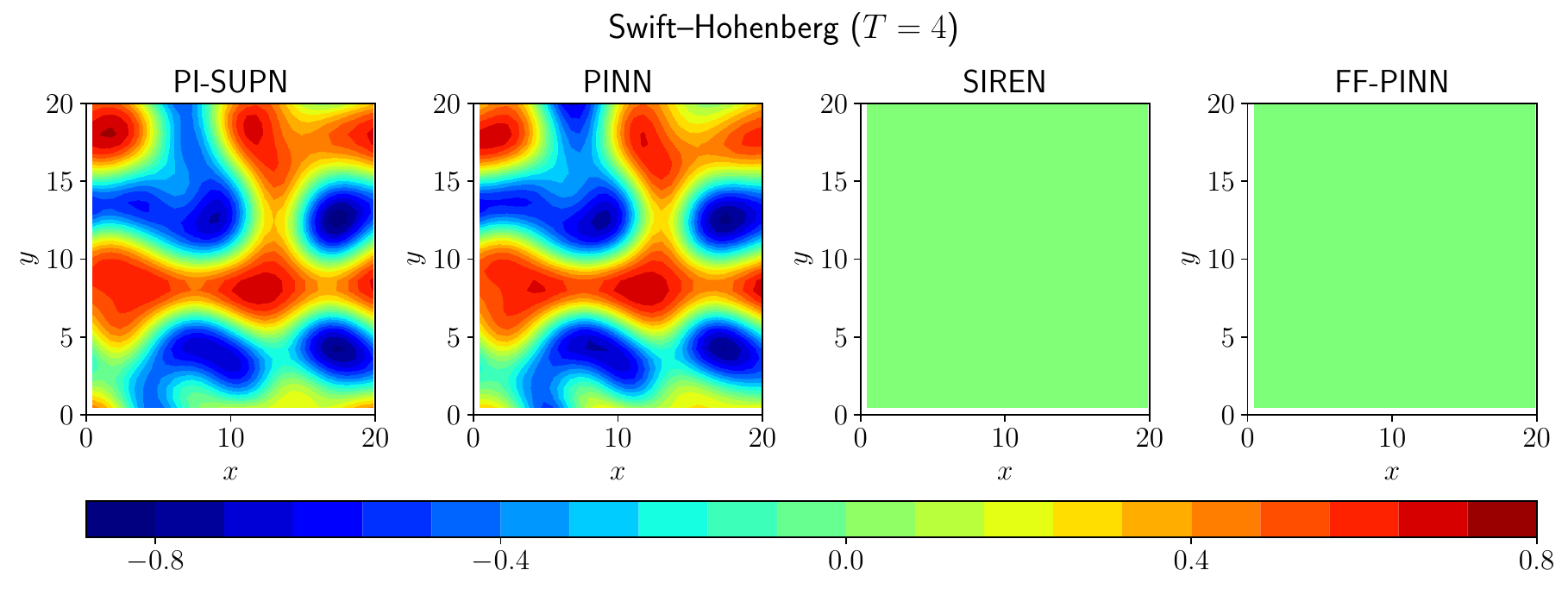}
    \caption{Model predictions at $T=4$ for Swift--Hohenberg (\Cref{sec:sh}), using FP64 and Adam + SS-BFGS. The reference solution is in \Cref{fig:sh_ref}.}
    \label{fig:sh_t_4_preds}
\end{figure}

To select the model for the final rollout, we perform an ablation study over width, depth/degree, precision, and optimizer for predicting the state at only $T=1$, shown in \Cref{fig:sh_t_1}. The models in \Cref{fig:sh_t_4_preds} are at the smallest model to attain within 10\% of the minimum error. We make this choice to avoid overstating the model size required to reach a given error level. \Cref{tbl:sh_t_4_err} shows the parameter counts and quantitative errors at $t=4$, corresponding to \Cref{fig:sh_rollout_error}.

\begin{table}[!htbp]
    \centering
    \begin{tabular}{l|c||c|c|c|c}
        \toprule
        & Method & Num. params. & Adam + SS-BFGS & Adam + SOAP & Adam only \\ \hline
        \multirow{4}{*}{\rotatebox[origin=c]{90}{FP64}} & PI-SUPN (ours) & \textbf{440} & $\mathbf{2.3 \times 10^{-1}}$ & $\mathbf{2.7 \times 10^{-1}}$ & $\mathbf{3.3 \times 10^{-1}}$ \\
        & PINN & 475 & $2.5 \times 10^{-1}$ & $3.4 \times 10^{-1}$ & $4.6 \times 10^{-1}$ \\
        & SIREN & 475 & $1.0 \times 10^{0}$ & $1.0 \times 10^{0}$ & $1.0 \times 10^{0}$ \\
        & FF-PINN & 585 & $1.0 \times 10^{0}$ & $1.0 \times 10^{0}$ & $1.0 \times 10^{0}$ \\ \hline
        \multirow{4}{*}{\rotatebox[origin=c]{90}{FP32}} & PI-SUPN (ours) & \textbf{440} & $\mathbf{2.3 \times 10^{-1}}$ & $\mathbf{2.6 \times 10^{-1}}$ & $\mathbf{3.3 \times 10^{-1}}$ \\
        & PINN & 475 & $2.5 \times 10^{-1}$ & $3.4 \times 10^{-1}$ & $4.5 \times 10^{-1}$ \\
        & SIREN & 475 & $1.3 \times 10^{0}$ & $1.0 \times 10^{0}$ & $1.2 \times 10^{0}$ \\
        & FF-PINN & 1197 & $1.0 \times 10^{0}$ & $1.0 \times 10^{0}$ & $1.0 \times 10^{0}$ \\
        \bottomrule
    \end{tabular}
    \caption{Prediction error at $t=4$ for various combinations of precision and optimizer in Swift--Hohenberg (\Cref{sec:sh}).}
    \label{tbl:sh_t_4_err}
\end{table}

The predicted solutions at $T=4$ are shown in \Cref{fig:sh_t_4_preds}. PI-SUPNs and PINNs obtain a similar shape at $T=4$ to the reference solution (\Cref{fig:sh_ref}). Even though approximation errors are close, \Cref{tbl:sh_t_4_runtime} shows that PINNs take approximately $1.5\times$ longer than PI-SUPNs to train, controlling for similar accuracy and model size. We attribute this difference to PINNs using automatic differentiation while PI-SUPNs use the three-term recurrence relation precomputed at the collocation points.

\begin{table}[!htbp]
    \centering 
    \begin{tabular}{l|c||c|c|c}
        \toprule
        & Method & Adam + SS-BFGS & Adam + SOAP & Adam only \\ \hline
        \multirow{4}{*}{\rotatebox[origin=c]{90}{FP64}} & PI-SUPN (ours) & $\mathbf{1.5 \times 10^{3}}$ & $\mathbf{1.0 \times 10^{3}}$ & $\mathbf{9.9 \times 10^{2}}$ \\
        & PINN & $2.3 \times 10^{3}$ & $1.5 \times 10^{3}$ & $1.5 \times 10^{3}$ \\
        & SIREN & $2.6 \times 10^{3}$ & $1.4 \times 10^{3}$ & $1.4 \times 10^{3}$ \\
        & FF-PINN & $3.8 \times 10^{3}$ & $3.7 \times 10^{3}$ & $4.0 \times 10^{3}$ \\ \hline
        \multirow{4}{*}{\rotatebox[origin=c]{90}{FP32}} & PI-SUPN (ours) & $\mathbf{8.1 \times 10^{2}}$ & $\mathbf{5.4 \times 10^{2}}$ & $\mathbf{5.3 \times 10^{2}}$ \\
        & PINN & $1.3 \times 10^{3}$ & $8.3 \times 10^{2}$ & $8.6 \times 10^{2}$ \\
        & SIREN & $9.5 \times 10^{2}$ & $8.1 \times 10^{2}$ & $8.6 \times 10^{2}$ \\
        & FF-PINN & $2.1 \times 10^{3}$ & $1.9 \times 10^{3}$ & $2.1 \times 10^{3}$ \\
        \bottomrule
    \end{tabular}
    \caption{Total training time (sec) using the models in \Cref{tbl:sh_t_4_err} for Swift--Hohenberg (\Cref{sec:sh}).}
    \label{tbl:sh_t_4_runtime}
\end{table}

\FloatBarrier
\subsection{Pedagogical One-Dimensional Examples}

\FloatBarrier
\subsubsection{High-Frequency} \label{sec:first-order-1d}

We first consider the following ODE problem \cite{moseley2023finite}:
\begin{align}
\frac{\d u}{\d x} &= 25 \pi \, \cos(25 \pi x), \; x\in(0,1] \\
u(0) &= 0 \nonumber
\end{align}
with exact solution $u(x) = \sin(25 \pi x)$. 

\rmk{Although this problem has high-frequency oscillations, its solution can be {\em exactly} represented by a Fourier Feature PINN or a SIREN PINN, which makes it useful as test case.}

\begin{figure}[!htbp]
    \centering
    \includegraphics[width=0.5\textwidth]{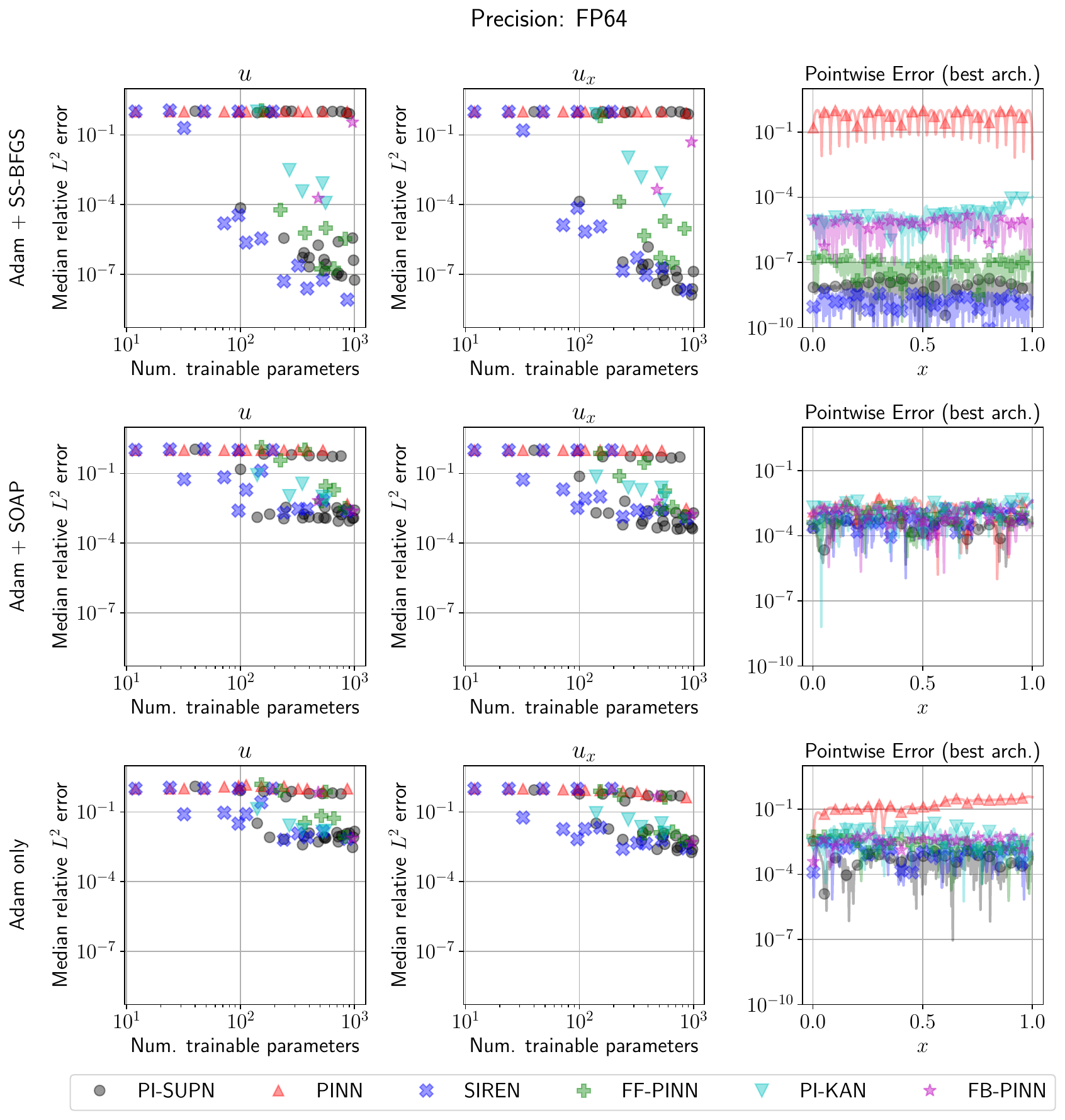}
    \caption{Ablation study using FP64 (expanded from \Cref{tbl:first-order-1d}). Left: Median error in $u(x)$ versus number of parameters. Center: Median error in $u'(x)$ versus number of parameters. Right: Pointwise prediction error.  Results are for the first-order 1D problem (\Cref{sec:first-order-1d}).}
    \label{fig:first-order-1d-convergence-fp64}
\par\medskip

    \includegraphics[width=0.5\textwidth]{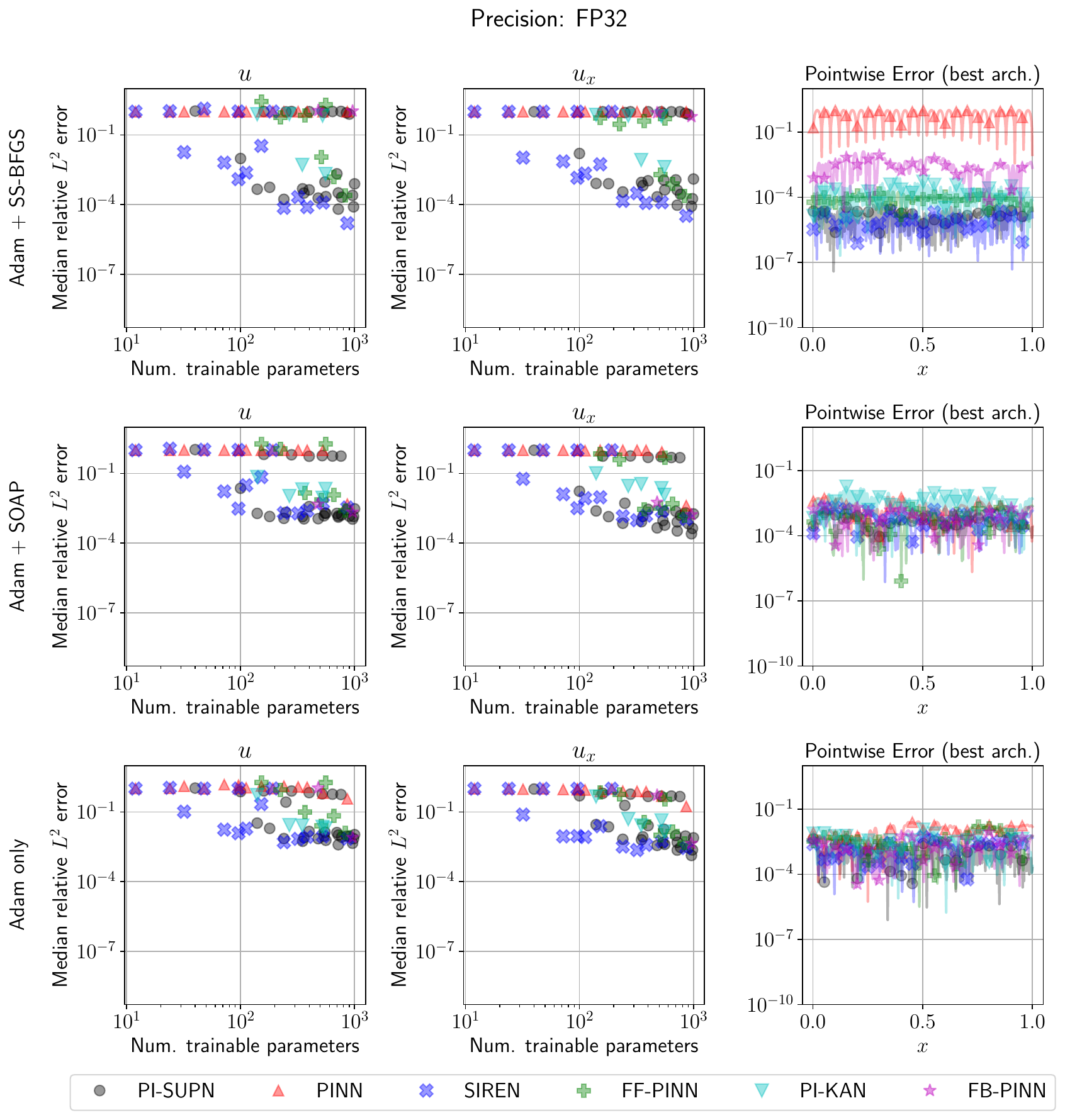}
    \caption{Ablation study using FP32 (expanded from \Cref{tbl:first-order-1d}). Left: Median error in $u(x)$ versus number of parameters. Center: Median error in $u'(x)$ versus number of parameters. Right: Pointwise prediction error.  Results are for the first-order 1D problem (\Cref{sec:first-order-1d}).}
    \label{fig:first-order-1d-convergence-fp32}
\end{figure}

We show the performance results in \Crefrange{tbl:first-order-1d}{tbl:first-order-1d-test-vs-train} and \Crefrange{fig:first-order-1d-convergence-fp64}{fig:first-order-1d-convergence-fp32}.  With SS-BFGS, SIRENs perform the best, as expected. Interestingly, Fourier-feature PINNs do not empirically obtain a higher accuracy than SUPNs here. \Cref{tbl:first-order-1d-test-vs-train} shows that computing solution errors at the collocation points generally leads to an underestimation of solution error at unseen locations. However, here it does not change the relative ranking of method performance.

\begin{table}[!htbp]
	\centering
    \small
	\begin{tabular}{l|l||c|c||c|c||c|c}
        \toprule
        & \multirow{2}{*}{Model} & \multicolumn{2}{c||}{Adam + SS-BFGS} & \multicolumn{2}{c||}{Adam + SOAP} & \multicolumn{2}{c}{Adam only} \\ \cline{3-8}
         & & $u$ & $u_x$ & $u$ & $u_x$ & $u$ & $u_x$ \\ \hline
        \multirow{6}{*}{\rotatebox[origin=c]{90}{FP64}} & PI-SUPN & ${6.2 \times 10^{-8}}$ & ${3.3 \times 10^{-8}}$ & $\bm{6.3 \times 10^{-4}}$ & $\bm{2.6 \times 10^{-4}}$ & $\bm{3.0 \times 10^{-3}}$ & $\bm{1.7 \times 10^{-3}}$ \\
        & PINN & $1.0 \times 10^{0}$ & $8.6 \times 10^{-1}$ & $4.9 \times 10^{-3}$ & $3.1 \times 10^{-3}$ & $7.8 \times 10^{-1}$ & $4.2 \times 10^{-1}$  \\
        & SIREN & $\bm{8.0 \times 10^{-9}}$ & $\bm{2.0 \times 10^{-8}}$ & ${2.2 \times 10^{-3}}$ & ${1.1 \times 10^{-3}}$ & ${6.4 \times 10^{-3}}$ & ${2.5 \times 10^{-3}}$ \\
        & FF-PINN & $1.7 \times 10^{-7}$ & $3.2 \times 10^{-7}$ & $3.3 \times 10^{-3}$ & $1.5 \times 10^{-3}$ & $8.4 \times 10^{-3}$ & $5.8 \times 10^{-3}$ \\
        & PI-KAN & $1.2 \times 10^{-4}$ & $1.6 \times 10^{-4}$ & $7.1 \times 10^{-3}$ & $8.6 \times 10^{-3}$ & $1.4 \times 10^{-2}$ & $1.2 \times 10^{-2}$  \\
        & FB-PINN & $1.9 \times 10^{-4}$ & $4.5 \times 10^{-4}$ & $2.4 \times 10^{-3}$ & $1.7 \times 10^{-3}$ & $7.4 \times 10^{-3}$ & $4.9 \times 10^{-3}$  \\ \hline 
        \multirow{6}{*}{\rotatebox[origin=c]{90}{FP32}} & PI-SUPN & ${5.5 \times 10^{-5}}$ & ${8.3 \times 10^{-5}}$ & $\bm{7.5 \times 10^{-4}}$ & $\bm{2.5 \times 10^{-4}}$ & $\bm{3.7 \times 10^{-3}}$ & $\bm{1.3 \times 10^{-3}}$ \\
        & PINN & $1.0 \times 10^{0}$ & $9.9 \times 10^{-1}$ & $4.8 \times 10^{-3}$ & $4.1 \times 10^{-3}$ & $3.7 \times 10^{-1}$ & $1.7 \times 10^{-1}$ \\
        & SIREN & $\bm{1.6 \times 10^{-5}}$ & $\bm{3.3 \times 10^{-5}}$ & ${1.9 \times 10^{-3}}$ & ${9.6 \times 10^{-4}}$ & ${5.0 \times 10^{-3}}$ & ${2.3 \times 10^{-3}}$ \\
        & FF-PINN & $2.4 \times 10^{-4}$ & $2.4 \times 10^{-4}$ & $2.6 \times 10^{-3}$ & $1.5 \times 10^{-3}$ & $1.2 \times 10^{-2}$ & $4.5 \times 10^{-3}$ \\
        & PI-KAN & $2.1 \times 10^{-3}$ & $4.1 \times 10^{-3}$ & $8.9 \times 10^{-3}$ & $1.2 \times 10^{-2}$ & $2.0 \times 10^{-2}$ & $1.4 \times 10^{-2}$ \\
        & FB-PINN & $1.1 \times 10^{0}$ & $6.2 \times 10^{-1}$ & $2.1 \times 10^{-3}$ & $1.7 \times 10^{-3}$ & $8.0 \times 10^{-3}$ & $4.4 \times 10^{-3}$ \\
	\bottomrule
	\end{tabular}
    \caption{Median relative $L^2$ error on the test set (over five initializations), using the best architectures from \Crefrange{fig:first-order-1d-convergence-fp64}{fig:first-order-1d-convergence-fp32}. Lowest error is in \textbf{bold}. Results are for the first-order 1D problem (\Cref{sec:first-order-1d}).}
    \label{tbl:first-order-1d}
\end{table}

\begin{table}[!htbp]
	\centering
	\begin{tabular}{l||c|c||c|c}
        \toprule
        \multirow{2}{*}{Model} & \multicolumn{2}{c||}{Rel. $L^2$ error ($\bm{x}_r$)} & \multicolumn{2}{c}{Rel. $L^2$ error ($\bm{x}_{\mathcal{I}}$)} \\ \cline{2-5}
        & $u$ & $u_x$ & $u$ & $u_x$ \\ \hline
        SUPN & $5.1 \times 10^{-8}$ & $1.3 \times 10^{-8}$ & $1.1 \times 10^{-8}$ & $9.5 \times 10^{-9}$ \\
        PINN & $1.0 \times 10^{0}$ & $8.6 \times 10^{-1}$ & $1.0 \times 10^{0}$ & $5.7 \times 10^{-1}$ \\
        SIREN PINN & $8.0 \times 10^{-9}$ & $2.0 \times 10^{-8}$ & $3.2 \times 10^{-9}$ & $1.3 \times 10^{-8}$ \\
        FF-PINN & $1.7 \times 10^{-7}$ & $3.1 \times 10^{-7}$ & $5.7 \times 10^{-8}$ & $2.7 \times 10^{-7}$ \\
        PI-KAN & $1.2 \times 10^{-4}$ & $1.6 \times 10^{-4}$ & $4.7 \times 10^{-5}$ & $8.9 \times 10^{-5}$ \\
        FB-PINN & $1.9 \times 10^{-4}$ & $4.4 \times 10^{-4}$ & $6.3 \times 10^{-5}$ & $1.7 \times 10^{-4}$ \\
        \bottomrule
    \end{tabular}
    \caption{Comparison of median relative $L^2$ errors on the test and training sets for Adam + SS-BFGS and FP64 at the best-performing architecture. Results are for the first-order 1D problem (\Cref{sec:first-order-1d}).}
    \label{tbl:first-order-1d-test-vs-train}
\end{table}

\FloatBarrier
\subsubsection{High-Order} \label{sec:fourth-order-1d}

Consider the following ODE problem
\begin{align}
\frac{\d^4 u}{\d x^4} &= (3\pi)^4 \cos(3\pi x), \qquad x\in(0,1] \\
u(0) &= 0, \;u'(0) = 1, \;u''(0) = - 9\pi^2, \; u'''(0)=6 \nonumber
\end{align}
with exact solution $u(x) = \cos( 3 \pi x) + x^3 + x -1$.

\begin{figure}[!htbp]
    \centering
    \includegraphics[width=0.5\textwidth]{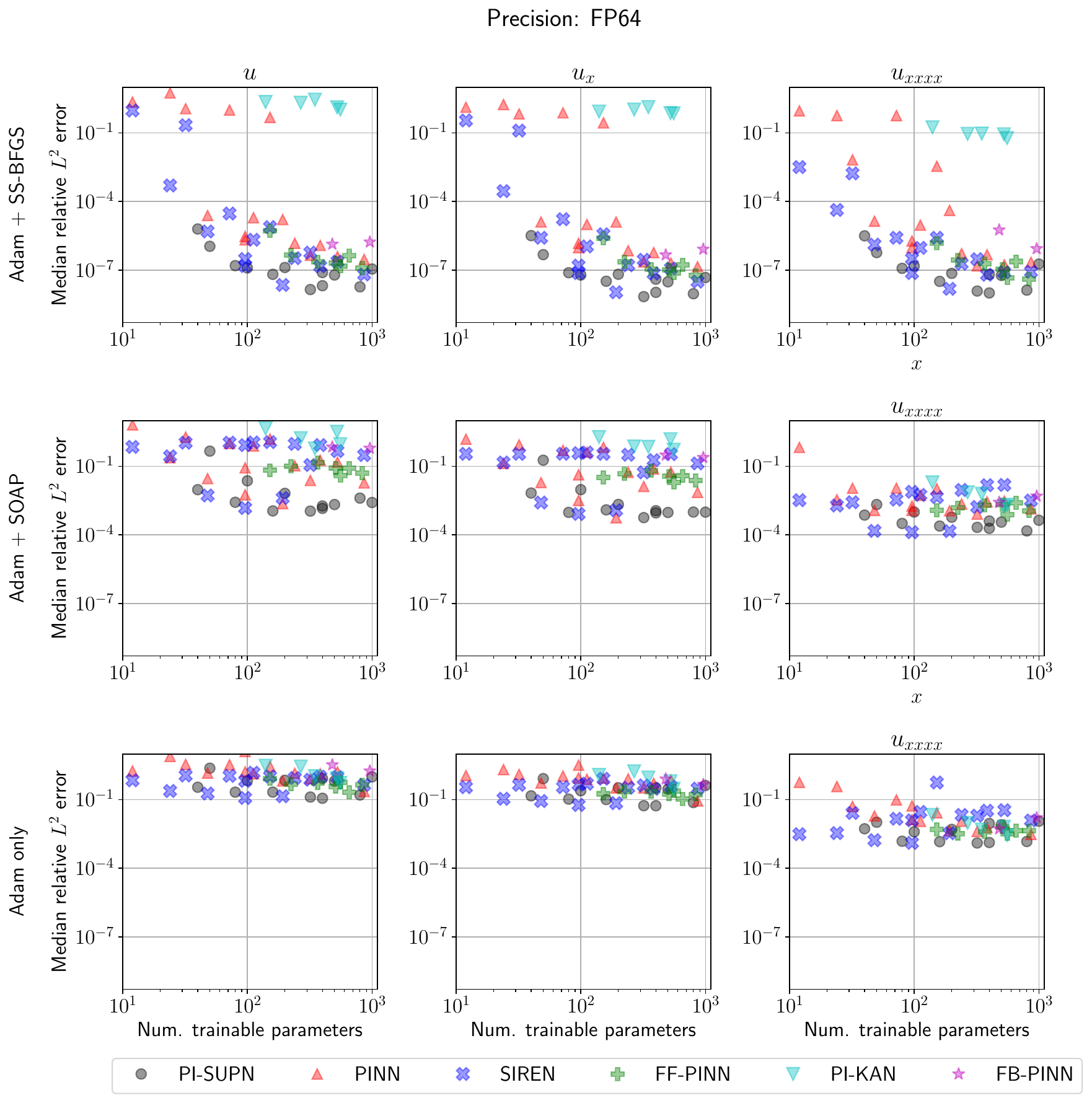}
    \caption{Ablations for FP64, showing median relative $L^2$ error (over five network initializations) versus number of parameters for $u$, $u_x$, and $u_{xxxx}$. Results are for the fourth-order 1D problem (\Cref{sec:fourth-order-1d}).}
    \label{fig:fourth-order-1d-fp64}
\par\medskip

    \includegraphics[width=0.5\textwidth]{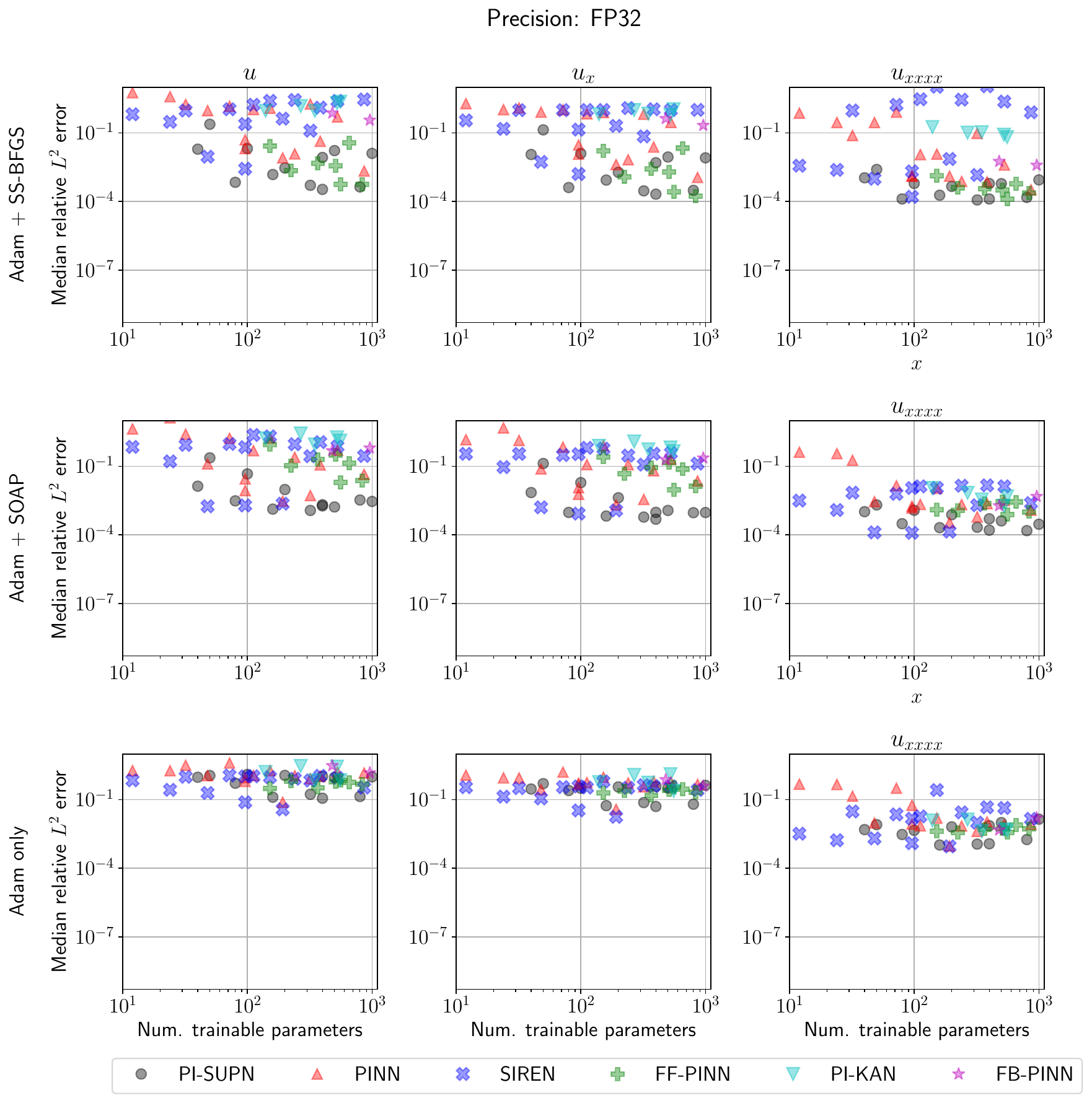}
    \caption{Ablations for FP32, median relative $L^2$ error (over five network initializations) versus number of parameters for $u$, $u_x$, and $u_{xxxx}$. Results are for the fourth-order 1D problem (\Cref{sec:fourth-order-1d}).}
    \label{fig:fourth-order-1d-fp32}
\end{figure}

We show the performance results in \Crefrange{tbl:fourth-order-1d}{tbl:fourth-order-1d-runtime} and \Crefrange{fig:fourth-order-1d-fp64}{fig:fourth-order-1d-fp32}. The polynomial terms in the solution cause the solution to be outside the exact approximation space of SIRENs. PI-SUPNs achieve the lowest error under FP64 and
Adam + SS-BFGS, while the relative rankings vary
across other optimizer and precision configurations.

\begin{table}[!htbp]
	\centering
	\begin{tabular}[c]{l | l || c | c | c }
        \toprule
        & Model & Adam + SS-BFGS & Adam + SOAP & Adam only \\ \hline
        \multirow{6}{*}{\rotatebox[origin=c]{90}{FP64}} & PI-SUPN & $\bm{1.4 \times 10^{-8}}$ & $\bm{1.1 \times 10^{-3}}$ & $\bm{1.1 \times 10^{-1}}$ \\
        & PINN & $2.9 \times 10^{-7}$ & $2.3 \times 10^{-3}$ & $2.3 \times 10^{-1}$ \\
        & SIREN & ${2.2 \times 10^{-8}}$ & ${1.5 \times 10^{-3}}$ & ${1.2 \times 10^{-1}}$ \\
        & FF-PINN & $1.2 \times 10^{-7}$ & $4.0 \times 10^{-2}$ & $2.1 \times 10^{-1}$ \\
        & PI-KAN & $1.2 \times 10^{0}$ & $7.0 \times 10^{-1}$ & $5.5 \times 10^{-1}$ \\
        & FB-PINN & $1.7 \times 10^{-6}$ & $6.1 \times 10^{-1}$ & $1.8 \times 10^{0}$ \\ \hline 
        \multirow{6}{*}{\rotatebox[origin=c]{90}{FP32}} & PI-SUPN & $\bm{3.4 \times 10^{-4}}$ & $\bm{1.3 \times 10^{-3}}$ & $\bm{1.2 \times 10^{-1}}$ \\
        & PINN & $2.1 \times 10^{-3}$ & $3.0 \times 10^{-3}$ & $7.7 \times 10^{-2}$ \\
        & SIREN & $2.7 \times 10^{-3}$ & ${1.8 \times 10^{-3}}$ & $3.8 \times 10^{-2}$ \\
        & FF-PINN & ${5.6 \times 10^{-4}}$ & $2.0 \times 10^{-2}$ & ${3.2 \times 10^{-1}}$ \\
        & PI-KAN & $9.2 \times 10^{-1}$ & $8.9 \times 10^{-1}$ & $7.0 \times 10^{-1}$ \\
        & FB-PINN & $3.6 \times 10^{-1}$ & $4.8 \times 10^{-1}$ & $1.5 \times 10^{0}$ \\
	\bottomrule
	\end{tabular}
    \caption{Median relative $W^{4,2}$ (Sobolev) error on the validation set (over five initializations), using the best architecture from \Crefrange{fig:fourth-order-1d-fp64}{fig:fourth-order-1d-fp32}. Lowest error is in \textbf{bold}. Results are for the fourth-order 1D problem (\Cref{sec:fourth-order-1d}).} 
    \label{tbl:fourth-order-1d}
\end{table}

\begin{table}[!htbp]
    \centering
    \begin{tabular}[c]{l|l||c|c|c|c|c}
        \toprule
        & Model & $u$ & $u_x$ & $u_{xx}$ & $u_{xxx}$ & $u_{xxxx}$ \\ \hline
        \multirow{6}{*}{\rotatebox[origin=c]{90}{FP64}} & PI-SUPN & $\bm{1.4 \times 10^{-8}}$ & $6.8 \times 10^{-9}$ & $1.9 \times 10^{-9}$ & $9.2 \times 10^{-10}$ & $9.9 \times 10^{-9}$ \\
        & PINN & $\bm{2.9 \times 10^{-7}}$ & $1.4 \times 10^{-7}$ & $3.3 \times 10^{-8}$ & $2.0 \times 10^{-8}$ & $1.5 \times 10^{-7}$ \\
        & SIREN & $\bm{2.2 \times 10^{-8}}$ & $1.1 \times 10^{-8}$ & $3.0 \times 10^{-9}$ & $4.0 \times 10^{-9}$ & $1.5 \times 10^{-8}$ \\
        & FF-PINN & $\bm{1.2 \times 10^{-7}}$ & $5.9 \times 10^{-8}$ & $1.5 \times 10^{-8}$ & $3.7 \times 10^{-9}$ & $4.0 \times 10^{-8}$ \\
        & PI-KAN & $\bm{1.2 \times 10^{0}}$ & $7.1 \times 10^{-1}$ & $3.0 \times 10^{-1}$ & $3.7 \times 10^{-1}$ & $5.9 \times 10^{-2}$ \\
        & FB-PINN & $\bm{1.7 \times 10^{-6}}$ & $4.6 \times 10^{-7}$ & $1.6 \times 10^{-7}$ & $6.5 \times 10^{-8}$ & $8.5 \times 10^{-7}$ \\ \hline 
        \multirow{6}{*}{\rotatebox[origin=c]{90}{FP32}} & PI-SUPN & $\bm{3.4 \times 10^{-4}}$ & $2.1 \times 10^{-4}$ & $7.0 \times 10^{-5}$ & $3.0 \times 10^{-5}$ & $1.2 \times 10^{-4}$ \\
        & PINN & $\bm{2.1 \times 10^{-3}}$ & $1.1 \times 10^{-3}$ & $3.1 \times 10^{-4}$ & $7.3 \times 10^{-5}$ & $3.2 \times 10^{-4}$ \\
        & SIREN & $\bm{2.7 \times 10^{-3}}$ & $1.6 \times 10^{-3}$ & $5.0 \times 10^{-4}$ & $1.2 \times 10^{-4}$ & $1.6 \times 10^{-4}$ \\
        & FF-PINN & $\bm{5.6 \times 10^{-4}}$ & $1.7 \times 10^{-4}$ & $3.6 \times 10^{-5}$ & $2.7 \times 10^{-5}$ & $1.3 \times 10^{-4}$ \\
        & PI-KAN & $\bm{9.2 \times 10^{-1}}$ & $6.1 \times 10^{-1}$ & $4.8 \times 10^{-1}$ & $4.2 \times 10^{-1}$ & $6.5 \times 10^{-2}$ \\
        & FB-PINN & $\bm{3.6 \times 10^{-1}}$ & $2.1 \times 10^{-1}$ & $6.8 \times 10^{-2}$ & $1.6 \times 10^{-2}$ & $3.9 \times 10^{-3}$ \\
        \bottomrule 
    \end{tabular}
    \caption{Separation of error from \Cref{tbl:fourth-order-1d} into each derivative, for Adam + SS-BFGS. {\bf Bold} indicates row-wise maximum, i.e. the corresponding entry in \Cref{tbl:fourth-order-1d}. Results are for the fourth-order 1D problem (\Cref{sec:fourth-order-1d}).}
    \label{tbl:fourth-order-1d-deriv}
\end{table}

\begin{table}[!htbp]
	\centering
    \begin{tabular}[c]{l||c|c|c|c|c|c}
    \toprule
    Model & SUPN & PINN & SIREN & FF-PINN & PI-KAN & FB-PINN \\ \hline
    Sec per epoch & ${1.5 \times 10^{-2}}$ & $5.6 \times 10^{-2} $ & $\bm{1.4 \times 10^{-2}}$ & $5.3 \times 10^{-2}$ & $7.3 \times 10^{-1}$ & $2.3 \times 10^{-1}$ \\
    Parameters & {320} & 864 & \textbf{192} & 832 & 556 & 960 \\
	\bottomrule
	\end{tabular}
    \caption{Parameter count and time of one SS-BFGS step, using the best-performing architecture from \Cref{fig:fourth-order-1d-fp64}. Results are for the fourth-order 1D problem (\Cref{sec:fourth-order-1d}).} 
    \label{tbl:fourth-order-1d-runtime}
\end{table}

\FloatBarrier
\section{Training Settings} \label{app:training}

All models were implemented in PyTorch and trained on NVIDIA A100 GPUs with 40GB HBM2 memory. Within each PDE problem, we use the same optimizer settings across all methods for Adam, SOAP, and SS-BFGS. We train for the number of epochs given in \Cref{tbl:num_epochs}.  We use~\cite{sitzmann2020implicit} for SIREN initialization, the Kaiming normal distribution for PI-SUPNs, and the Kaiming uniform distribution for everything else~\cite{he2015delving}.

\begin{table}[!htbp]
    \centering
    \begin{tabular}{c||c|c}
        \toprule
        PDE & Burn-in (Adam) & SS-BFGS/SOAP/Adam \\ \hline
        Helmholtz & 2000 & 23000 \\
        Allen--Cahn & 5000 & 20000 \\
        Euler--Bernoulli Beam & 10000 & 10000 \\
        Swift--Hohenberg & 4000 & 2000 \\
        Pedagogical 1D & 2000 & 28000 \\
        \bottomrule 
    \end{tabular}
    \caption{Number of epochs in training.}
    \label{tbl:num_epochs}
\end{table}

We use the Scimba defaults in SS-BFGS (\texttt{https://gitlab.com/scimba/scimba/-/tree/main}, commit hash \texttt{91ad86fb}), except for a gradient tolerance of $10^{-12}$. We use the default settings in SOAP (\texttt{https://github.com/nikhilvyas/SOAP}, commit hash \texttt{a1e5535}) except for a learning rate of $10^{-3}$ and preconditioning frequency of 2. For Adam, we use a learning rate of $10^{-3}$ for the 1D problems and $10^{-2}$ for everything else.